\documentclass{aa}  

\usepackage{graphicx}
\usepackage{booktabs}
\usepackage{placeins}
\usepackage{txfonts}
\usepackage[colorlinks=true,linkcolor=black,citecolor=blue,urlcolor=magenta]{hyperref}
\begin{document}

   \title{Vertical Structure of Protoplanetary Disks in Scattered Light: A large sample analysis}

   \author{J. Byrne
          \inst{1,2}
          \and
          C. Ginski\inst{1}
          \and
          R. F. van Capelleveen\inst{4}
          \and
          N. Fitzgerald\inst{1}
          \and
          A. Garufi\inst{3}
          \and
          C. Coyne\inst{1}
          \and
          C. Lawlor\inst{1}
          \and
          D. McLachlan\inst{1}
          }

   \institute{School of Natural Sciences, Center for Astronomy, University  of Galway, Galway, H91 CF50, Ireland
        \and
            School of Physics, Trinity College Dublin, The University of Dublin, College Green, Dublin 2, Ireland
         \and
             INAF, Istituto di Radioastronomia, Via Gobetti 101, I-40129, Bologna, Italy
        \and
            Leiden Observatory, Leiden University, Postbus 9513, 2300 RA Leiden, The Netherlands\\
             }

   \date{Received October 28, 2025; accepted March 3, 2026}
 
  \abstract 
   {High-resolution imaging in scattered light has revealed complex morphologies in protoplanetary/circumstellar disks. Measuring their vertical height is key to understanding disk structure, evolution, and the properties of embedded dust.}
   {This work aims to develop a robust methodology for fitting elliptical shapes to scattered light images of protoplanetary disks in order to extract vertical height profiles of the dust scattering surface ($\tau = 1$) across a large and morphologically diverse disk sample. The dataset comprises 92 unique near-infrared polarimetric images of individual systems obtained with VLT/SPHERE. The goal is to identify trends in vertical structure across morphologies and test for correlations with stellar mass, age, and disk dust mass. Using the height profiles, this work also investigates the implications of the constrained height for the masses of potential embedded planets.}
   {A structure extraction and ellipse fitting (SEEF) algorithm, is implemented using edge detection and Gaussian fitting to locate the structure within protoplanetary disks. Fitting ellipses to the structure reveals spatial offsets from the centre of the ellipse and the star, interpreted as vertical height assuming circular ring geometry. Disk inclination, position angle (PA), and aspect ratio ($h_{\tau=1}/r$) are also derived.}
   {The SEEF algorithm obtained successful vertical height measurements for 92 unique disks, revealing variations in height profiles consistent with flared disk geometries. Analysis of the full sample shows that the vertical height profile cannot be confidently described by a single power-law relation. Subdivision of the sample by disk morphology revealed no strong correlations within most categories, with the exception of extended disks ($r_{\mathrm{outer}} \geq 150$ au), which exhibited a strong correlation with a single power-law trend. Investigation into underlying disk properties revealed no correlation for its affect to the vertical height structure.}
   {This work presents a consistent methodology for measuring the vertical structure of circumstellar disks using ellipse fitting on scattered light images. While trends in height structure remain weakly correlated for the full sample and many disk morphologies, extended disks ($r_{\mathrm{outer}} \geq 150$ au) stand out as the only subgroup showing a clear power-law flaring trend. The lack of a strong correlation across other morphologies and with system properties like stellar or dust mass suggests that either differing disk morphologies exhibit different vertical height profiles or that another, unidentified factor is influencing the disk flaring.}


   \maketitle

\section{Introduction}

    \begin{figure*}[t]
        \centering
        \includegraphics[width=0.98\textwidth]{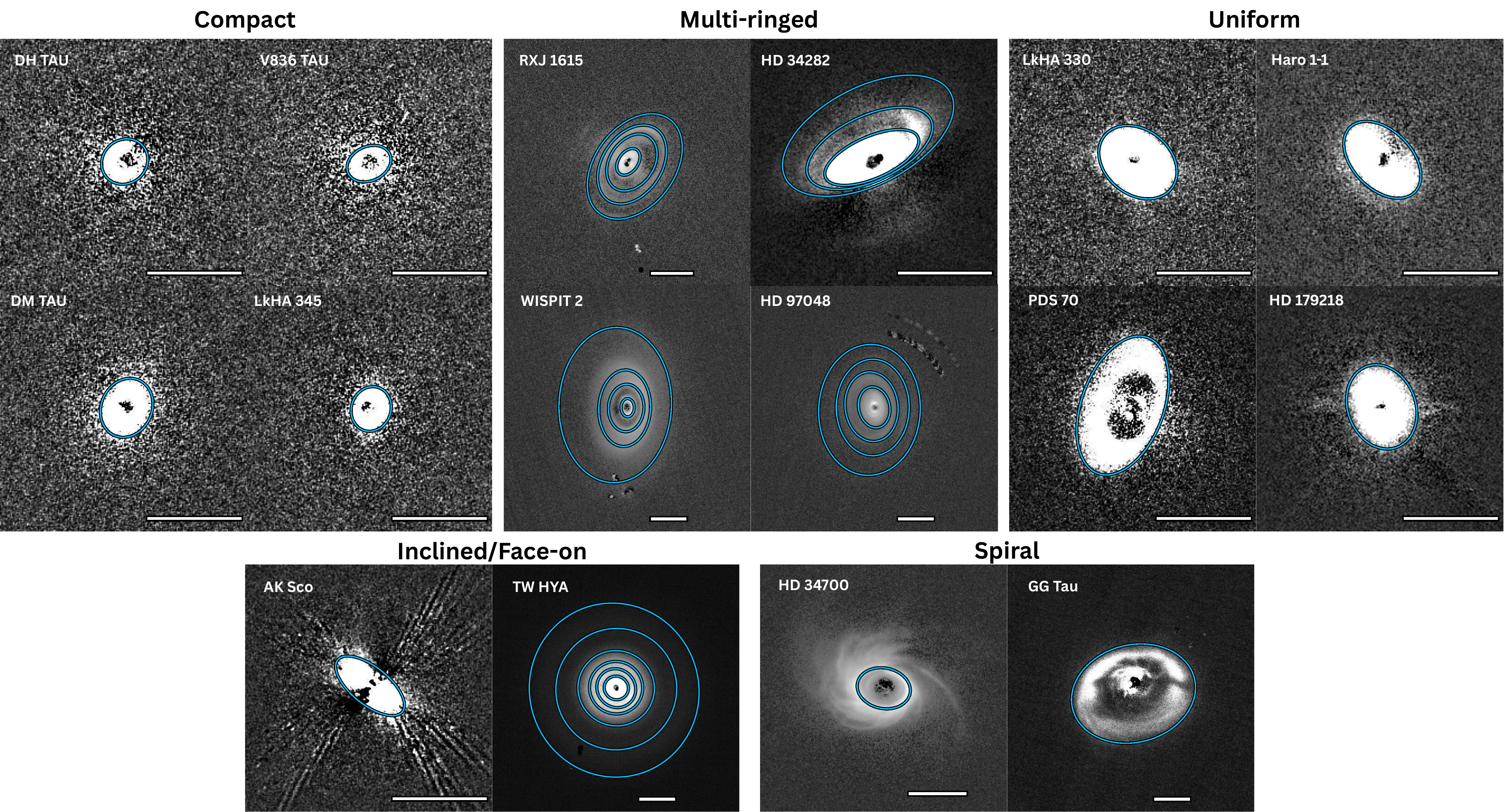}
        \caption{Gallery of representative disks from the sample, grouped by morphological geometry. Ellipse fits made in this study to the scattered-light disk structures are overplotted. The scale bar in each panel denotes 1 arcsecond.}
        \label{fig:gallery}
    \end{figure*}

   Circumstellar disks are the birthing grounds of planets. High spatial resolution observations taken in the past decade have revealed complex disk morphologies, such as multi-ringed disks, spirals and gaps, which are evidence of ongoing dynamical processes. Understanding the physical geometry of these disks is essential for constraining the disk evolution and the processes that lead to planet formation. However, it still remains unclear how the current observed planet population relates to the disk properties. In this context, understanding the vertical height structure of the dust is a necessary step toward linking observed disk features to the mechanisms that shape planetary systems. While several studies have measured the dust vertical height in individual disks or small samples \citep{de_Boer_2016,Ginski_2016,Avenhaus2018,Rich2021,Ginski2024}, a large-scale analysis is needed to identify broader trends.

   The first set of imaged protoplanetary disks mostly consisted of Herbig Ae/Be stars \citep[e.g.,][]{Grady1999,Fukagawa2006}, which are traditionally divided into two groups: Group I sources, whose spectral energy distributions show strong far-infrared excesses interpreted as flared disks with large inner cavities, and Group II sources, which display weaker far-infrared emission corresponding to flatter disks with small or no inner cavities \citep[e.g.,][]{Garufi2014,Honda2015}, leading to the long-standing study concerning the presence of a disk cavity and the disk height. Group II disks have been found to be inherently faint in scattered light \citep{Grady2005}, thus any large sample is systematically biased towards their exclusion, making a comparison in scattered light difficult. However, by investigating the dust vertical height structure for different categories of disk morphology it is possible to identify trends that may indicate the presence or absence of a cavity.

   Further, the disk height may be affected by other properties of the disk. The degree of disk flaring is expected to correlate with the disk’s total mass. The disk becomes more optically thick with an increase in total dust mass, therefore observations of the optically thick scattering layer trace higher layers within the disk. Therefore, a relationship between dust mass and the measured aspect ratio is physically motivated. Over time, the disk’s aspect ratio is expected to decrease as the system evolves. This reflects the physical processes that cause the disk to become geometrically thinner, such as radiative cooling, mass depletion due to accretion or expulsion from the system, and reduced vertical support from a lack of gas pressure, resulting in a smaller dust vertical height relative to radius. If this is correct, the aspect ratio should decline with increasing stellar age. Alternatively, the clearing of the cavity observed in older (transition) disks could enhance the irradiation of the outer disk, heating it more strongly and thereby increasing its flaring. Lastly, the vertical height of the dust may correlate with stellar mass (see Appendix \ref{StellarMassCalc} for dependency derivation), since the classical mass-luminosity relation (MLR) established for main-sequence stars is not expected to hold during the pre–main-sequence phase (e.g., \citet{Baraffe2002}) characteristic of protoplanetary disks. Further, \citet{Eker2018} propose that the MLR for main-sequence stars may be described by $L \propto M^\beta_*$ with differing values for $\beta$ depending on stellar mass, reflecting transitions in the dominant energy transport mechanism. Therefore, there may be a correlation between vertical height and stellar mass.

   In this study, we present a large-sample analysis of the vertical height structure of the optically thick dust scattering surface ($\tau = 1$) within 92 protoplanetary disks, comprised of polarimetric near-infrared (NIR) scattered-light observations. This sample represents a substantial fraction of all NIR scattered-light images of protoplanetary disks currently available.
   
   This paper is organised as follows. In Sect. \ref{Observations} details of the dataset used throughout this work are presented. Sect. \ref{Methods}, outlines the methods used to extract the disk vertical height structure. Finally, Sect. \ref{Results & Discussion} contains the results and discussion of the full dataset, two example individual systems, multi-ringed systems, an investigation into system properties that may affect the dust's vertical height structure and the calculation of planet masses opening gaps, a use case for the constraint of the dust vertical height structure.

\section{Observations}
\label{Observations}

    \begin{figure}[b!]
        \centering
        \includegraphics[width=1\linewidth]{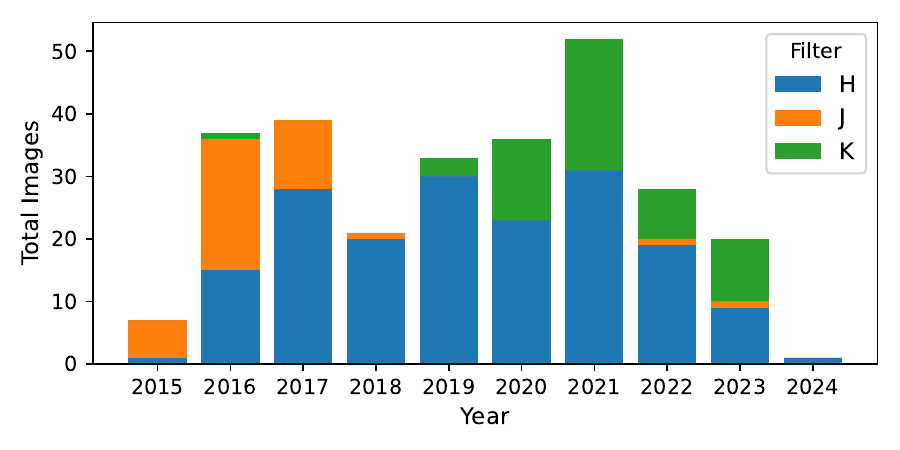}
        \caption{The distribution of year and filter within the initial dataset (totalling 294 images), showing a greater tendency towards H-band images.}
        \label{fig:ImageDistributionResults}
    \end{figure}

    \begin{figure*}[t]
    \centering
    \sidecaption
        \includegraphics[width=0.7\textwidth]{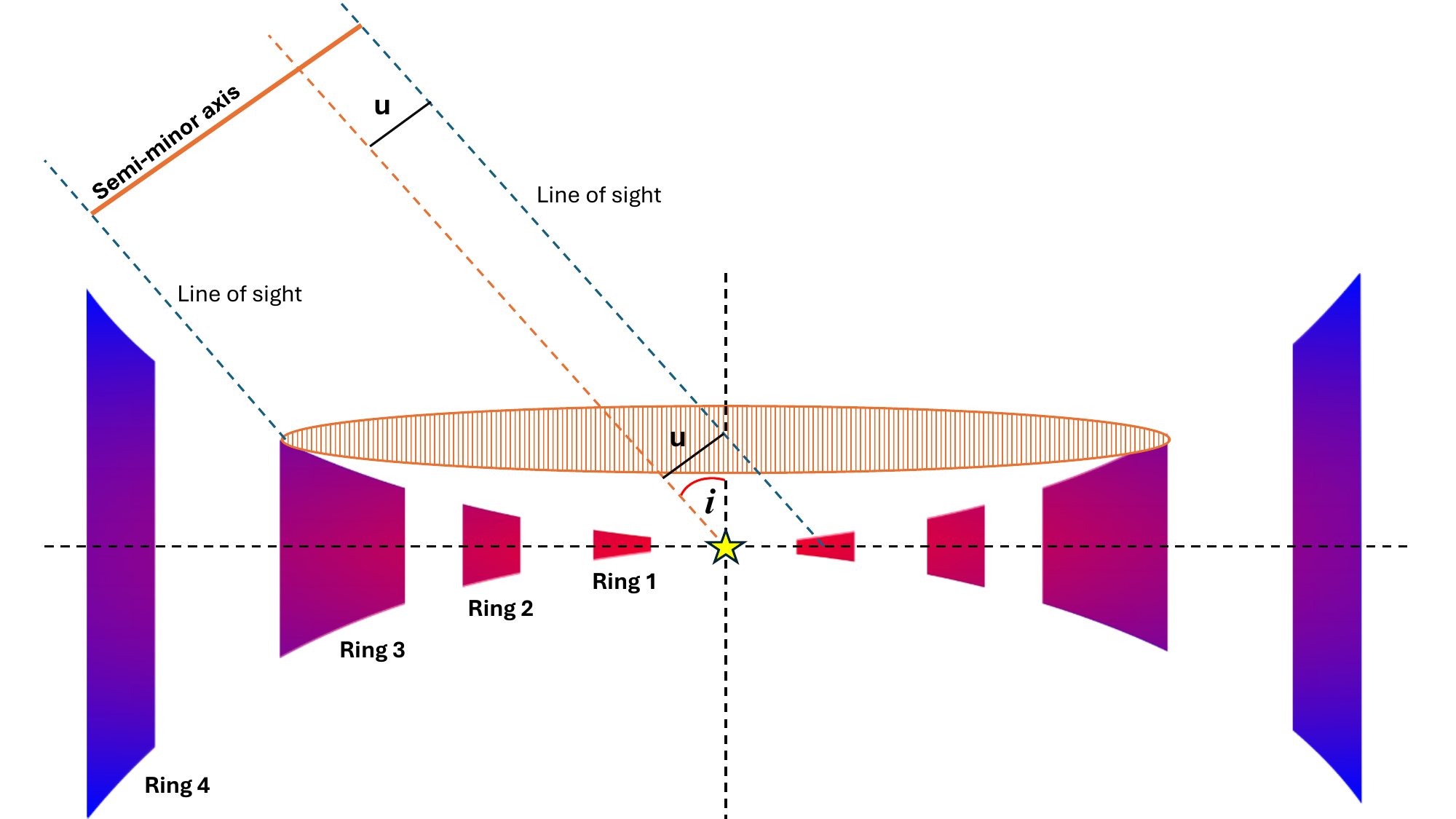}
        \caption{Schematic illustration of the projected geometry of an inclined multi-ringed circumstellar disk (this also holds for continuous disks). When viewed at an inclination angle $i$, the disk’s intrinsically (and assumed) circular structure appears elliptical due to projection. The offset $u$ between the star position and the centre of the ellipse arises from the projected height of the disk surface along the semi-minor axis. This geometric offset becomes observable in scattered light images and is used to infer disk height structure.}
        \label{fig:offset}
    \end{figure*}
   The dataset used in this study comprises the full set of protoplanetary disk observations available in the European Southern Observatory (ESO) archive from the Very Large Telescope using the Spectro-Polarimetric High-contrast Exoplanet REsearch instrument (VLT/SPHERE; \citealt{Beuzit2019}) as of September, 2024, representing a substantial fraction of all existing polarised, scattered-light images of protoplanetary disks to date. The dataset consists of 186 unique disks with 294 total images, observed with VLT/SPHERE using the IRDIS (Infra-Red Dual-beam Imager and Spectrograph; \citealt{Dohlen2008}) subsystem. The dual-beam polarimetric imaging (DPI; \citealt{Langlois2014}) mode was used in each case, measuring linear polarised light corresponding to the scattered light from the dust scattering surface. The dataset is comprised of J (1.1 - 1.4 $\mu$m), H (1.5 - 1.8 $\mu$m) and Ks (2.0 - 2.4 $\mu$m; hereafter K-band) band images taken between 2015 and 2024 (see Fig.~\ref{fig:ImageDistributionResults}). All data were reduced using IRDAP (IRDIS Data reduction for Accurate Polarimetry) which is described in detail in \citet{van_Holstein_2020}. The reduction includes a model-based correction for instrumental polarisation and crosstalk between optical components.

   To differentiate signal strength in the images and to evaluate the disks’ suitability for the structure extraction/ellipse fitting (SEEF) algorithm (see Sect. \ref{Methods}), they were classified as high-confidence, moderate-confidence or low-confidence cases. High-confidence disks required little to no user intervention, moderate-confidence disks presented challenges such as asymmetries or spiral features, and low-confidence disks were dominated by noise or exhibited insufficient disk signal for reliable detection. 71 disks were categorised as high-confidence, 21 as moderate-confidence and 92 as low-confidence. Thus, the SEEF algorithm was run for 92 disks total (high- and moderate-confidence disks), using 1 image per disk (as many disks had observations in two or more filters). The image with the least distortions within the field-of-view was chosen.

\section{Methods}
\label{Methods}
    
    Following \citet{de_Boer_2016}, to extract the vertical height profile of the optically thick dust scattering surface of the disk, ellipses are fit to the observed structures to measure the offset, $u$, between the centre of the fitted ellipse and the stellar position along the minor axis. Shown in Fig \ref{fig:offset}, the height of the scattering surface of the dust, where the disk becomes optically thick to the wavelength of light ($h_{\tau = 1}$, hereafter simply $h$), is calculated as $h = u/\sin (i)$, where $i$ is the inclination of the disk. The separation, inclination and position angle (PA) are also derived from the ellipse fit.

    The algorithm\footnote{The SEEF algorithm is public and available to download at https://github.com/JakeByrne7/SEEF} created to fit ellipses and measure the vertical height structure consists of two main stages: structure extraction and ellipse fitting (SEEF). In the first stage, positional information of the disk structure is extracted from the image (Sect. \ref{Structure Extraction}). In the second stage, elliptical models are fitted to this information to characterise the geometry (Sect. \ref{Ellipse Fitting}).

    \subsection{Pre-processing}
    \label{Pre-Processing}

    Pre-processing of the images is often required to increase the algorithm's capability of structure extraction. Firstly, the image is rotated to an optimal starting position for outlier removal (see Sect. \ref{outlierremoval}). Optionally, a high pass filter, specifically unsharp masking \citep{Malin1977}, may be applied to enhance structures. A smoothed representation was generated by convolving with a Gaussian kernel ($\sigma = 5$ pixels). The high-frequency component was obtained by subtracting the smoothed image from the original. This residual was then added back to the original image with a gain factor of 2. This step was used only for disks with high S/N ratios and relatively smooth edges (e.g, high-confidence images with edges of S/N > 5). For disks containing faint features, unsharp masking can alter the apparent structure due to the low contrast between signal and noise. The high-pass filter was therefore used primarily to aid edge detection, but not for Gaussian fitting, as it often disrupted the intrinsic shape of the signal profile.

    The second optional pre-processing tool is an interactive ellipse-based masking tool, developed in Python\footnote{Development of the tool was assisted by AI and is also publicly available at https://github.com/JakeByrne7/SEEF} to isolate specific regions of interest, specifically for use during structure extraction in this work. The tool allows the user to define elliptical regions directly onto images using mouse-driven selection, with support for three masking modes: masking the interior of a single ellipse ("inner"), the exterior ("outer"), or the annular region between two ellipses ("ring"). Users can toggle between modes, zoom in and out, interactively adjust the position and size of ellipses, and apply or reset masks in real time.  Once finalised, the tool outputs the masked image to the SEEF algorithm, effectively enabling precise control over which regions of the disk are included or excluded in subsequent measurements. The tool was used for structures within multi-ringed disks where signal from adjacent rings interfered with extracting the target structure or for images containing artifacts that required masking.

    \subsection{Structure Extraction}
    \label{Structure Extraction}

    \begin{figure*}[t]
        \center
        \includegraphics[width=0.98\textwidth]{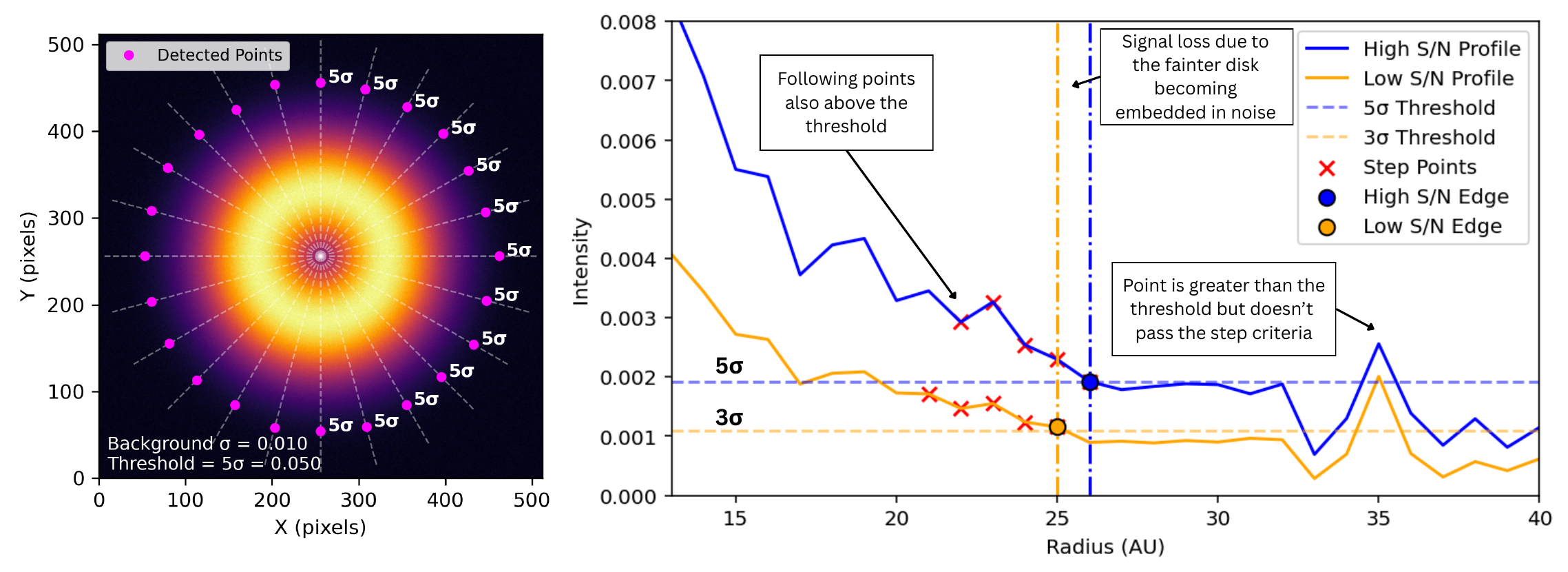}
        \caption{\textbf{Left:} An idealistic example of the edge detection method using a synthetic circumstellar disk with added Gaussian background noise. Radial cuts (dashed white lines) sample intensity profiles iteratively. Magenta dots mark detected disk edges where intensity first (measured inwards towards the star) exceeds a 5$\sigma$ noise primary threshold and passes the step function criteria. Edge labels are shown only on the right side of the disk (angles between 0$^\circ$ and 180$^\circ$ clockwise) for clarity. Background noise level ($\sigma$) and primary threshold (5$\sigma$) are annotated in the fig.\textbf{ Right:} Radial intensity profile extracted at 180$^\circ$ from the centre of GM Aur in H-band, comparing the real high-S/N case (blue) and a synthetically degraded low-S/N case (orange). Horizontal dashed lines indicate the 5$\sigma$ and 3$\sigma$ primary thresholds used for edge detection. Step points used to confirm edge detection are marked with red crosses, and the final detected edge locations are indicated by vertical dashed-dotted lines and filled circles. The low-S/N profile shows degraded detection performance and premature truncation due to the outer edge being buried in noise.}
        \label{fig:EdgeDetection}
    \end{figure*}

    Two structure-extraction methods are available for use within the SEEF algorithm, edge detection and Gaussian fitting, following and building on methods described by \citet{Ginski2024}.

    Edge detection is performed by placing a radial cut through the image to detect the outer disk edge. The most outward point along the radial cut with intensity above a user-defined threshold (hereafter called the primary threshold) is stored as a signal point (see Fig.~\ref{fig:EdgeDetection}). The direction of the radial cut is then rotated through the centre and the process repeated azimuthally in $1^\circ$ steps. Binning is applied based on the radial extent of the feature, where features at larger radii ($r > 75$~px / 0.92") are left unbinned to avoid undersampling. For features within approximately 100~px, Nyquist sampling is achieved, whereas extended features ($r > 100$~px / 1.22") are undersampled, resulting in the loss of minor fine structure along the disk edge. Nonetheless, this level of sampling is sufficient for accurately extracting the disk edge location. The primary threshold is defined relative to the background standard deviation ($\sigma$), with 5$\sigma$ used as a standard primary threshold for the majority of disks, with lower $\sigma$ primary thresholds used for faint structures when required.

    To address the issue of single-pixel high-intensity noise, a step-function was implemented. In this technique, the algorithm checks the next $n$ pixels along the radial direction, and if their intensities also exceed the set primary threshold, the point is marked as a valid signal point. The choice of $n$ is a user-defined parameter and dependent on the target structure: if set too high, narrow features may be overlooked or true signal points discarded due to individual noisy pixels, whereas setting it too low makes the method more susceptible to positive noise. Following \citet{Ginski2024}, the radial uncertainty ($\Delta r$) is calculated as:
    \begin{equation}
        \Delta r = \frac{3 \sigma}{m_{\mathrm{profile}}} 
    \end{equation}
    where $m_{\mathrm{profile}}$ is the slope of the step function's $n$ pixels calculated along the radial cut.

    The effect of the scattering phase function (i.e., a brighter forward-scattering and fainter back-scattering side of the disk) must be considered when using the edge detection technique. Using a fixed, user-defined azimuthal threshold on inclined disks can cause the algorithm to trace different physical regions of the disk surface due to strong azimuthal intensity variations. This represents a significant methodological caveat of the approach. The impact of the scattering phase function on this method is examined in detail by \citet{Ginski2024}, who apply the technique to two sets of radiative transfer models with differing degrees of flaring ($\alpha = 1.09$ and $\alpha = 1.30$), with both sets spanning a range of inclinations. By comparing the recovered inclination, position angle, and aspect ratio to the known model inputs, they show that the measured parameters remain consistent with the true values within the uncertainties. Notable caveats include a systematic offset of approximately 0.05 of the aspect ratio for the higher flaring case ($\alpha = 1.30$), where the measured values were systematically lower than the true model values. Disks with inclinations($i \ge 65^\circ$) were not included in this analysis and for lowly inclined disks ($i \le 10^\circ$), the uncertainties on PA and inclination were large, making the measurements unreliable. Further, as discussed in Appendix \ref{outlierremoval}, the SEEF algorithm permits manual exclusion of azimuthal bins; for inclined disks ($i \ge 45^\circ$), bins on the back-scattering side were frequently excluded to mitigate the effects of azimuthal intensity variations introduced by the scattering phase function.

    \begin{figure*}[t]
    \centering
    \sidecaption
        \includegraphics[width=0.70\textwidth]{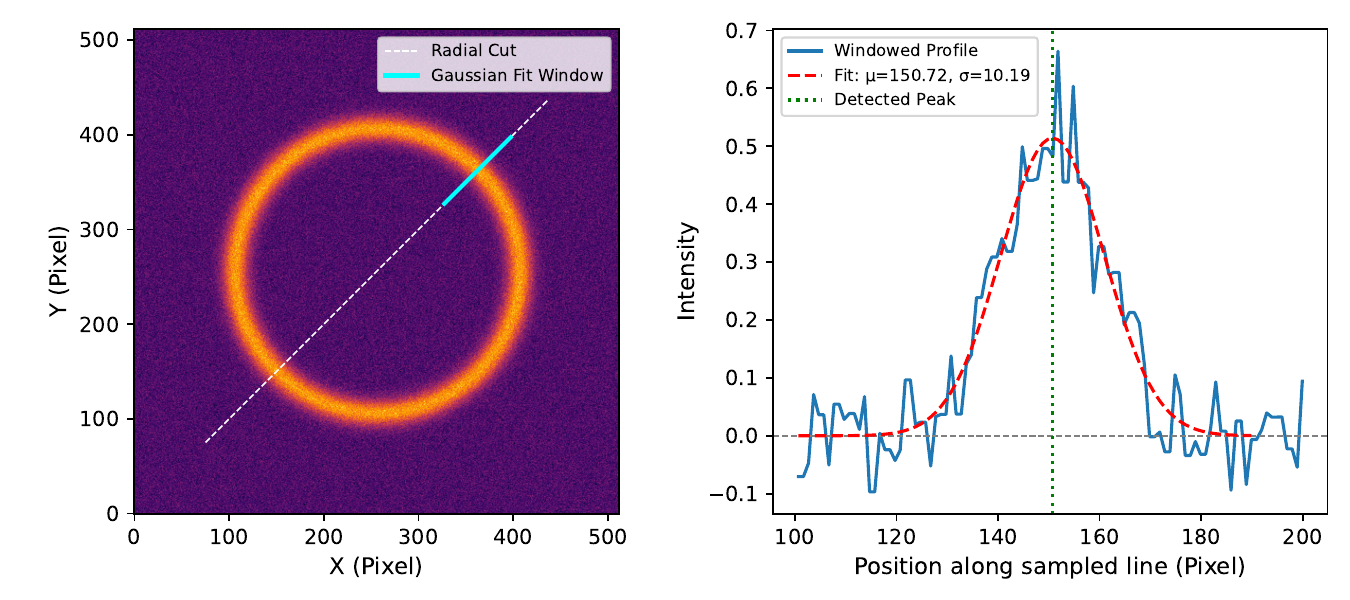}
        \caption{An idealistic example of Gaussian fitting. Shown is a synthetic circumstellar ring with added noise (\textbf{left}) showing the sampled radial cut (white) and the Gaussian fit window (cyan). The \textbf{right} panel displays the extracted intensity profile along the window and the Gaussian fit (red dashed line) modelling the ring edge, with the detected peak position indicated by the green vertical line. \textbf{Note:} here $\sigma = \Sigma$ in the main text.}
        \label{fig:GaussianDet}
    \end{figure*}

    The edge detection method is well-suited for identifying the boundaries of continuous disks and faint extended structures, where the target is resolved and exhibits a distinct edge as it detects the most outward signal and the user-defined threshold can be adjusted depending on the S/N of the target, with a higher threshold level for higher S/N disks. However, for inclined disks, the method is sensitive to azimuthal intensity variations introduced by the scattering phase function, which can cause the detected edge to correspond to different physical regions of the disk surface in forward- and back-scattering directions. It is also not effective for detecting radially-thin ring-like features. 
    
    Gaussian fitting was implemented as a complementary technique. This technique leverages both the signal strength and the shape of the intensity profile as criteria for extracting the ring's location. The implementation involved sampling intensity profiles along a window of a radial cut and fitting a Gaussian function of the form:

    \begin{equation}
    I(x) = A \exp\left(-\frac{(x - \mu)^2}{2\Sigma^2}\right),
    \end{equation}
    
    where \( A \) is the amplitude, \( \mu \) the centre (peak position), and \( \Sigma \) the standard deviation (related to the profile width). A sliding window technique was used to scan through each profile in fixed-size segments.

    To filter meaningful detections, the algorithm enforced several constraints: the peak amplitude \( A \) had to exceed a user-defined minimum (generally a more liberal 3$\sigma$ of the background), and the fitted width \( \Sigma \) had to fall within reasonable physical bounds, based on the width of the ring structure. For a ring structure approximately 10 pixels wide, this corresponds to \( \Sigma \approx 4.25 \) pixels, since the full width at half maximum (FWHM) is related to $\Sigma$ by FWHM $= 2.355 \cdot \Sigma$. Therefore, the bounds for acceptable fits were chosen to bracket this value, typically between 2 and 7 pixels, to allow for moderate variation while excluding unrealistically narrow or broad features. This helped suppress noise and false, narrow peaks. On fulfilling this criteria, the peak of the Gaussian is stored as the detected point. Following \citet{ginski2025}, an uncertainty estimate was derived as the FWHM divided by the S/N, where the S/N was defined as \( A \) divided by the standard deviation of the background.

    Unlike edge detection, which identifies the location of sharp intensity changes to define the outer edge of a ring, Gaussian fitting targets the centre of the intensity profile, effectively detecting the brightest point in the emission rather than the boundary. While this makes it conceptually different from edge detection, the resulting measurements of vertical height (see Table \ref{AllTable}) are functionally equivalent. Gaussian fitting also mitigates the impact of the scattering phase function compared to edge detection by providing a centroid estimate rather than relying on raw intensity. However, the stricter criteria and the additional complexity of fitting often result in fewer valid detections overall, especially in low S/N regions or when the profile deviates from a clean Gaussian shape.

    Following the extraction of data points, a series of outlier removal steps were performed on the data and are explained in detail in Appendix~\ref{outlierremoval}.
 
    \subsection{Ellipse Fitting}
    \label{Ellipse Fitting}

    Two forms of ellipse fitting are employed: a constrained and a free version of the least-squares method.
    
    The free ellipse fitting algorithm employed is the Direct Least Squares (DLS) method, introduced by \citet{Fitzgibbon1995ABG}. The method minimises the algebraic distance between the set of points derived from structure extraction and a fitted conic, subject to a quadratic constraint (4$ac - b^2$ = 1) to ensure the fitted conic is an ellipse, avoiding degenerate solutions. It reduces to solving a generalised eigenvalue problem, making it fast and numerically stable.

    It is important to note that the DLS fitting algorithm inherently biases the solution toward smaller ellipses due to its use of algebraic distance as the objective function, rather than the true geometric (orthogonal) distance between data points and the ellipse. While algebraic distance offers computational efficiency and enables a closed-form solution under certain constraints, it does not uniformly penalise deviations across the ellipse and disproportionately favours solutions with smaller semi-axes. This issue has been discussed in detail by \citet{Kanatani1994}, who demonstrated that minimising algebraic distance introduces a systematic bias that is not easily corrected.

    \begin{figure*}[t]
        \centering
        \includegraphics[width=0.98\textwidth]{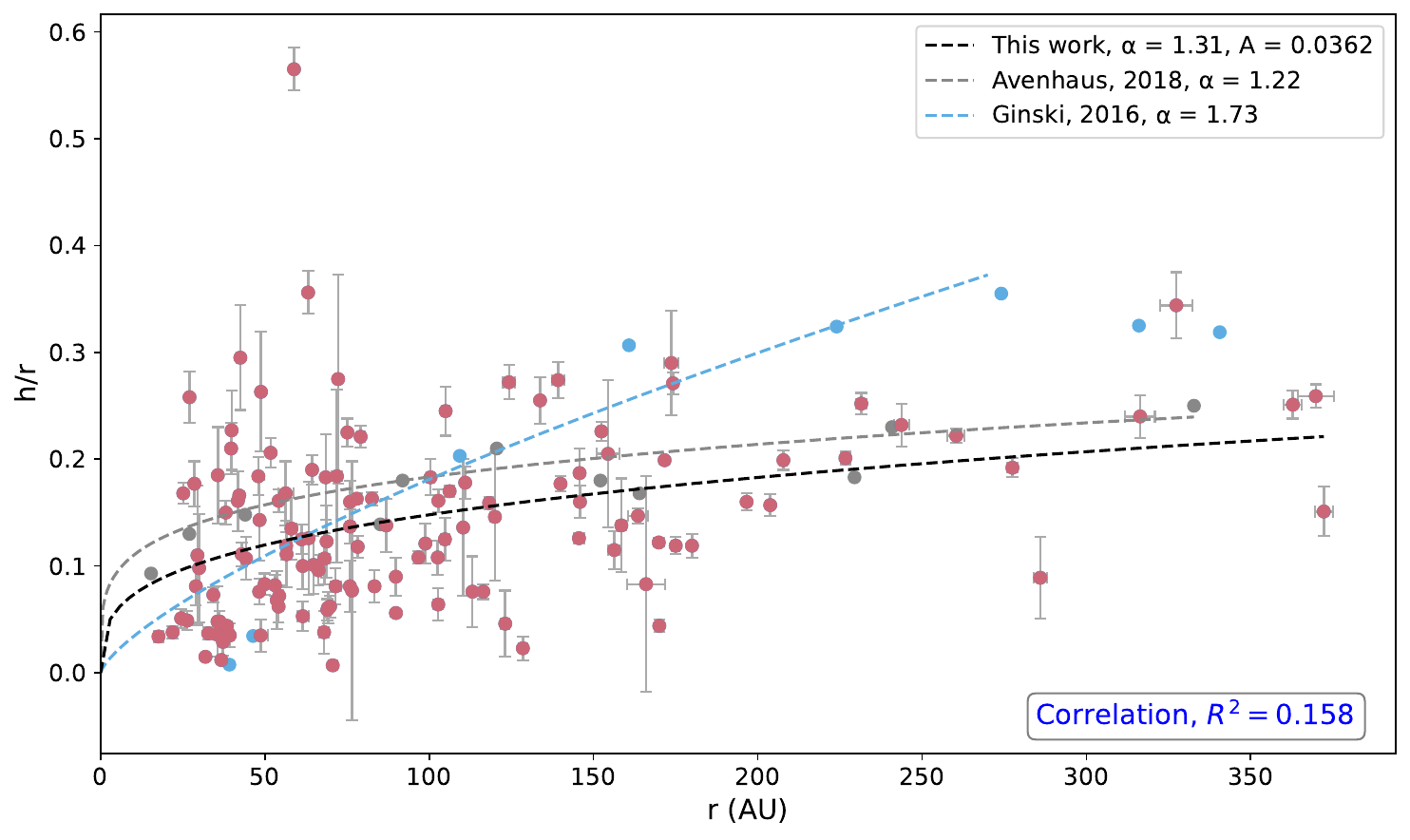}
        \caption{Aspect ratio ($h/r$) versus separation (r) for the structures within the 92 disks, with 133 measurements (rose). A power-law model of the form $h/r = A  \left(\frac{r}{1\mathrm{au}}\right)^{\alpha - 1}$ was fit to the data using least squares (black). ($R^2$) is shown in the bottom-right corner. For comparison, power-law fits from \citet{Avenhaus2018} and \citet{Ginski_2016} are overlaid along with their corresponding data points in grey and light blue respectively. }
        \label{fig:hrvsrall}
    \end{figure*}

    The constrained fit is built upon the SciPy \texttt{least\_squares} function \citep{2020SciPy-NMeth} which utilises the Levenberg-Marquardt (LM) algorithm \citep{1978LNM...630..105M}. The constrained version of the algorithm incorporates prior measurements of PA and inclination by initialising the solution with values obtained from previous ALMA observations, thereby anchoring the fit in physically motivated bounds and improving convergence in cases with noisy edge data, particularly when only a limited arc of the structure is visible. This technique enforces the conic constraint to ensure that the solution represents a valid ellipse that realistically defines the unseen structure. Although PA and inclination could be fixed, they were instead provided as initial guesses for the least squares function rather than hard constraints. This allows the fit to remain flexible and account for uncertainties in those parameters. This form of the ellipse fit is suitable for cases with limited edge coverage (see Appendix~\ref{Coverage and PA convention} for further details on the coverage required for each ellipse fit form) and to address the free fit's tendency towards the minimum solution. Therefore, for systems with prior available measurements of inclination and PA from ALMA-based observations, the constrained ellipse fit was always used (Table \ref{AllTable} indicates which form was used for each system).

    Uncertainty estimation for the fitted ellipse parameters differs depending on whether constraints are applied. In the constrained case, errors are derived analytically from the Jacobian matrix returned by the LM optimisation. For the free fit, a non-parametric bootstrap method is used. This involves repeatedly resampling the edge point data with replacement and re-fitting the ellipse in each iteration (100,000 iterations used in this work). The distribution of recovered parameters across bootstrap trials provides empirical estimates of the standard deviation, which are taken as the uncertainty on each fitted parameter. Most disks are adequately sampled near the Nyquist limit; however, extended or low-contrast disk structures are often undersampled in specific azimuthal directions. As a result, uncertainty estimates from both Jacobian-based least-squares fitting and non-parametric bootstrap resampling may be dominated by well-sampled regions of the disk and may underestimate errors on global shape parameters.

    The offset (used to calculate the height, $h$), inclination, radial separation (semi-major axis, $r$) and PA (see Appendix~\ref{Coverage and PA convention} for PA convention used within this work) are measured directly from the ellipse fit. The aspect ratio ($h/r$) is then calculated as a measure of the disk flaring, which can be described by a power law, 

    \begin{equation}
        \label{eqn:h/r}
        \frac{h}{r} = \frac{h_0}{r_0}\left(\frac{r}{r_0}\right)^{\alpha-1},
    \end{equation}
    
    where $h_0$ describes the height at radius $r_0$, and $\alpha$ is the flaring index, which quantifies the degree of flaring found within the disk.

\section{Results and Discussion}
\label{Results & Discussion}

    The properties measured for each of the 92 disks can be found in Table \ref{AllTable} and the images with ellipses overlaid in Appendix \ref{AllDiskFigure}. 
    
    A series of plots are presented within this section, showing $h$ versus $r$ and $h/r$ versus $r$, illustrating the vertical structure of disks. The coefficient of determination, \( R^2 \), is used to evaluate how well the fitted models reproduce the observed data. A value of \( R^2 = 1 \) indicates a perfect fit, while \( R^2 = 0 \) implies that the model explains none of the variance in the data. Negative values of \( R^2 \) can occur when the model fits the data worse than a horizontal line at the mean, meaning the sum of squared residuals from the model exceeds the variance of the data. For vertical height, a power-law of the form  $h = A \cdot \left(\frac{r}{1\mathrm{au}}\right)^{\alpha}$ is fit to the dataset, where $\alpha$ is the flaring index and $A$ is the scaling parameter, with \( R^2 \) used to quantify how well this curve captures the observed spread. For the aspect ratio, \( R^2 \) is computed using a fitted power-law model of the form \( h/r = A \cdot \left(\frac{r}{1\mathrm{au}}\right)^{\alpha - 1} \). In both cases, higher \( R^2 \) values indicate stronger agreement between model predictions and data, serving as a comparative metric.
    
    The $R^2$ value for aspect ratio versus radius is lower than that for height versus radius because dividing by $r$ suppresses large-scale radial trends and reduces the variance of the dependent variable, inherently limiting the fraction of variance that can be explained by a power-law model in radius. However, by removing global radial scaling, $h/r$ suppresses the dominance of large-radius points in unnormalised fits. Since this analysis compares many different disks, parameters $A$ and $\alpha$ from $h/r$ fits are adopted as less biased, more consistent measures of disk flaring. Accordingly, the $R^2$ from aspect ratio space is reported as the meaningful measure of conformity.

\subsection{The Full Sample}
\label{Global Results}

    To evaluate trends across the dataset, the full sample of disks is analysed collectively. By combining the measured vertical height structures from all systems, patterns arising from disk properties or morphology can be identified.

    The plot of aspect ratio vs. the separation is shown in Fig.~\ref{fig:hrvsrall}. The resulting value of $R^2 = 0.158$ indicates that a single power-law model explains only a small fraction of the variance in the observed aspect ratios. This suggests a weak correlation between $h/r$ and radius across the sample, and implies that disk flaring may not follow a homogenous power-law relation for all disk morphologies. Given the low overall correlation, reflected in the weak $R^2$ value of 0.158, a single flaring (this plot yields $\alpha = 1.31$) relation for the full-disk sample may not adequately capture the diversity in disk structures. The disk flaring is compared against two previous studies with height measurements; \citet{Ginski_2016} measured the rings and gaps of a single multi-ringed disk HD~97048, while \citet{Avenhaus2018} measured the heights of structures within five disks around T Tauri stars. Our measured trend closely follows that of \citet{Avenhaus2018} who find a flaring index $\alpha = 1.22$, but with a lower overall mean height. Compared to \citet{Ginski_2016}, a stronger flaring is found in this work at small separations ($r \lesssim 60$~au), while their larger flaring index ($\alpha = 1.73$) leads to greater flaring at larger separations.

    To investigate whether underlying trends exist within subsets of the population, the disks were classified into distinct morphological categories based on observed structure. The categorisation morphological criteria was as follows: those with a total radius exceeding 150 au were classified as extended; disks with an inclination greater than $70^\circ$ were labelled as inclined; disks with an inclination lower than $15^\circ$ were labelled as face-on; systems exhibiting visible spiral arms were identified as spiral disks; and those with a radius smaller than 50~au were designated as compact. Disks not meeting the criteria for the specific groups were assigned to the uniform category. These disks exhibit a mostly smooth, axisymmetric morphology with no distinct features that would allow further classification.

    Three groups that posed a challenge to the fitting routine due to their morphology and the effect of the scattering phase-function as detailed in Sect. \ref{Structure Extraction} were identified: spiral, inclined and face-on. This is primarily due to their distinct structural characteristics and the limitations of the fitting procedure. Spiral disks often exhibit non-axisymmetric features, such as arms or warps, that deviate from the smooth elliptical morphology of a typical protoplanetary disk, introducing challenges that can compromise structure detection. Both highly and lowly inclined disks suffer from a common sensitivity issue: any slight deviation in the fitted ellipse parameters leads to a disproportionately large change in the inferred offset, owing to the strong dependence of the offset term on the inclination angle. For highly inclined systems, this effect is amplified because projection effects stretch the disk along the major axis, making the semi-minor axis, and thus the height measurement, extremely sensitive to even small fitting errors. For lowly inclined systems, the offset is intrinsically small because $\sin~i \approx 0$.  The offset is therefore nearly undetectable in the image, and any deviation in the measurement will cause a large relative change in the height measurement.

    \begin{figure}
        \centering
        \includegraphics[width=1\linewidth]{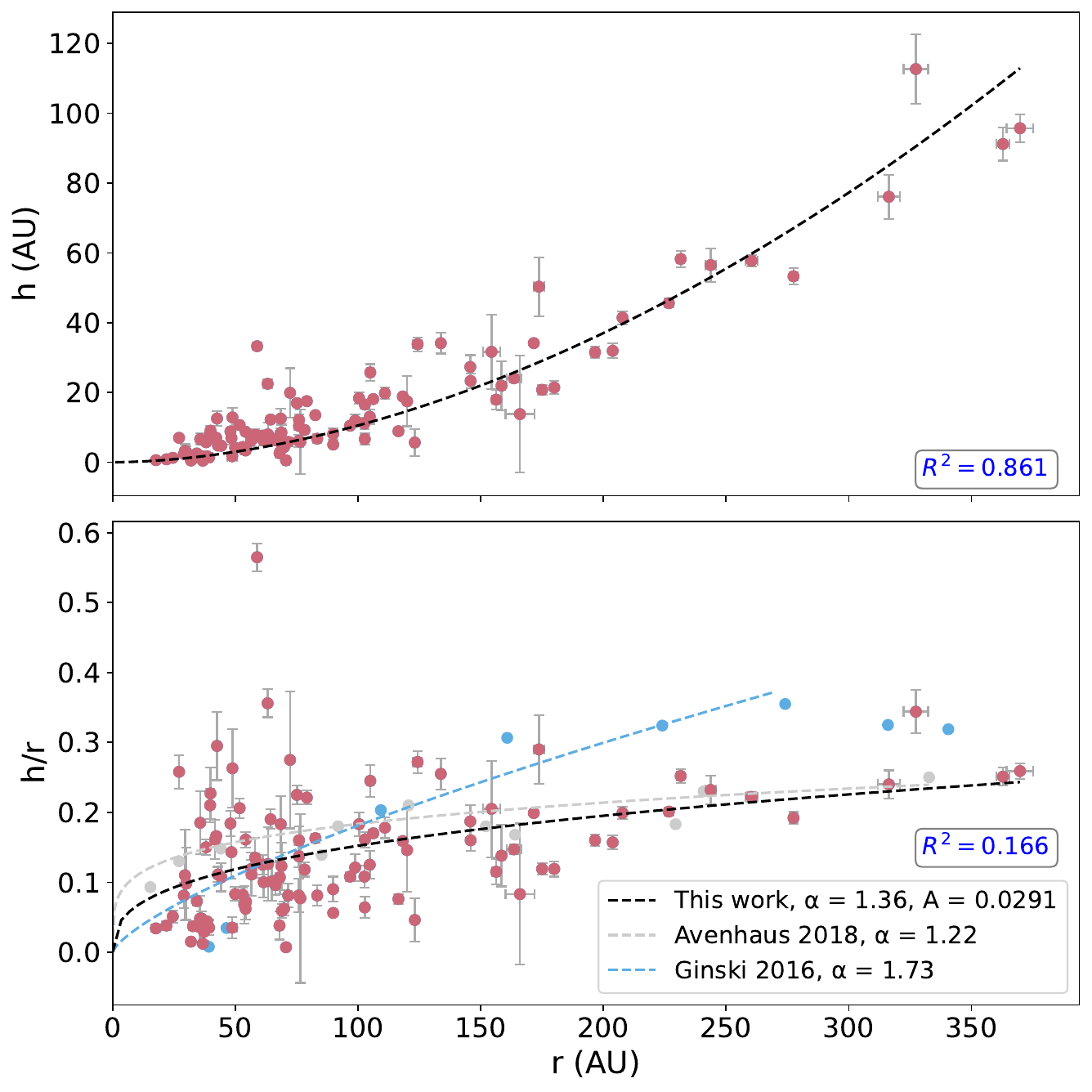}
        \caption{Two-panel visualisation of disk vertical height ($h$) and aspect ratio ($h/r$) as a function of separation ($r$) for the restricted dataset, with spiral and highly or lowly inclined disks excluded. Leaving extended, compact, and uniform disk types totalling 107 measurements.  $R^2$ is shown in the bottom right of each plot.
        \textbf{Top:} Power-law fit to $h$ vs. $r$ shows an increasing vertical structure consistent with flaring.
        \textbf{Bottom:} $h/r$ vs. $r$ with power-law fits overlaid for this work (black), and comparison to \citet{Avenhaus2018} and \citet{Ginski_2016} in grey and light blue respectively.}
        \label{fig:cut3}
    \end{figure}

    \begin{figure}[h]
        \centering
        \includegraphics[width=1\linewidth]{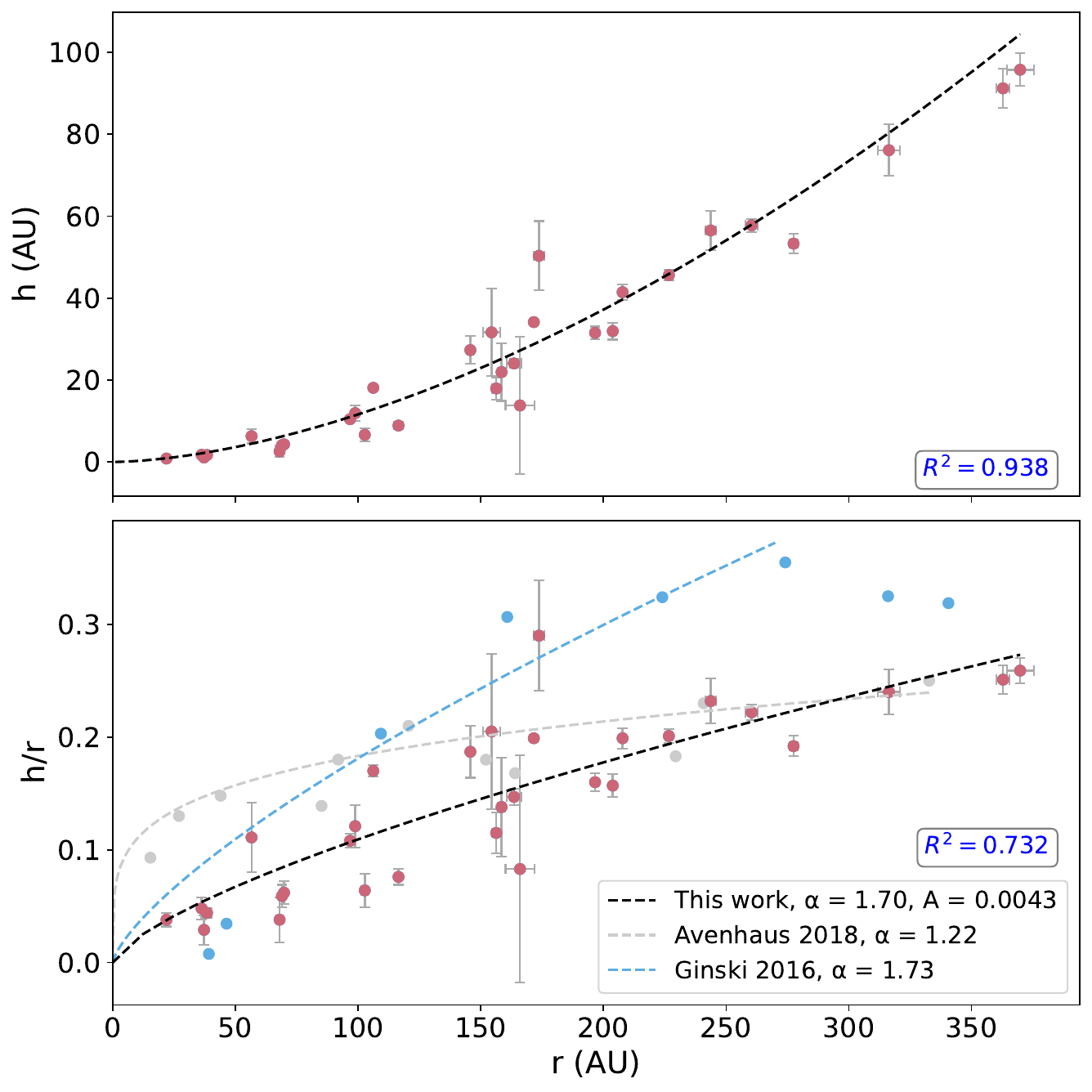}
        \caption{Two-panel visualisation of disk vertical height ($h$) and aspect ratio ($h/r$) as a function of separation ($r$) for extended disks only (with a total extent of $>$150~au). $R^2$ is shown the in bottom right of each plot.
        \textbf{Top:} Power-law fit to $h$ vs. $r$ shows an increasing vertical structure consistent with flaring.
        \textbf{Bottom:} $h/r$ vs. $r$ with power-law fits overlaid for this work (black), and comparison to \citet{Avenhaus2018} and \citet{Ginski_2016} shown in grey and light blue respectively. }
        \label{fig:extended3}
    \end{figure}
    
    Excluding the aforementioned categories and plotting the remaining groups together as a restricted dataset (totalling 79 unique disks and 107 measurements), namely the extended, compact, and uniform disks, is shown in Fig.~\ref{fig:cut3}. The restriction results in a negligible increase in the coefficient of determination ($R^2 = 0.166$). This increase suggests that even when the difficult-to-fit disks are removed, the scatter is still present. An $R^2$ value of 0.166 implies that 16.6\% of the variance in the observed aspect ratios can be explained by the model and while this is still a relatively weak correlation, it indicates that the power-law trend is not entirely absent but that a single flaring prescription is insufficient to capture the full diversity of vertical structures in the entire population.

    When plotting the same graphs for extended disks only (see Fig.~\ref{fig:extended3}), $R^2$ increases significantly across both representations ($R^2_{\mathrm{height}} = 0.938$ and $R^2_{\mathrm{h/r}} = 0.732$). This stronger correlation indicates that extended disks more closely follow the expected power-law flaring behaviour, consistent with theoretical models \citep{Chiang1997,Kenyon1987}. This subset thus provides the clearest empirical support for a flared disk geometry which may be confidently described by a single power-law, among all morphological classes analysed. 

    The power-law fit to the extended disk category gives a flaring index of 1.70 with $A = 0.0043$, which is in close agreement with the flaring index of 1.73 reported by \citet{Ginski_2016}, though the scaling factor is slightly lower than their value of $A = 0.0064$. This suggests that while the overall radial dependence of the vertical height in the extended disk subsample is consistent with previous findings, the vertical heights are systematically smaller. This discrepancy in $A$ could reflect differences in the sample selection (this work includes 11 unique disks, whereas \citet{Ginski_2016} analysed only a single disk) or fitting methodology. Nonetheless, the agreement in the flaring exponent reinforces the presence of a broadly similar underlying disk geometry. Compared to the full sample shown in Fig.~\ref{fig:hrvsrall}, which follows a trend similar to \citet{Avenhaus2018}, the extended disk sample's trend more closely resembles the results of \citet{Ginski_2016}, whose measurements were based a single extended disk. Considering \citet{Avenhaus2018} had a more varied dataset (containing disks that are classified as extended and uniform), this may have introduced a similar trend to that found in Fig. \ref{fig:hrvsrall}.

    The vertical scale height of mm-sized dust grains has been measured in multiple ALMA studies \citep[e.g.,][]{Villenave2020, Doi2021, Jiang2025}, consistently yielding values of a few au at radii equal to 100~au, corresponding to an aspect ratio of approximately 0.05 or lower. Independent JWST/MIRI observations \citep{Villenave2024, Tazaki2025}, which estimate the widths of dark lanes in edge-on disks, are consistent with these values and find similar aspect ratios. Equivalently, this work finds the aspect ratio to be approximately 0.1 at 100~au, suggesting the micron-sized dust is better coupled to the gas and are distributed at higher layers when compared to mm-sized dust which tends more towards the midplane of the disk. This further demonstrates that infrared observations predominantly trace the optically thick upper layers of the disk rather than the bulk dust mass.

\subsection{Multi-Ringed Systems and WISPIT 2}
\label{Multi-Ringed Systems & WISPIT 2}

    This section explores the entire multi-ringed population and WISPIT~2 \citep[TYC 5709-354-1;][]{VanCapelleveen2025}, a multi-ringed disk recently resolved in the NIR for the first time. 

\begin{figure*}[t]
        \centering
        \includegraphics[width=1\linewidth]{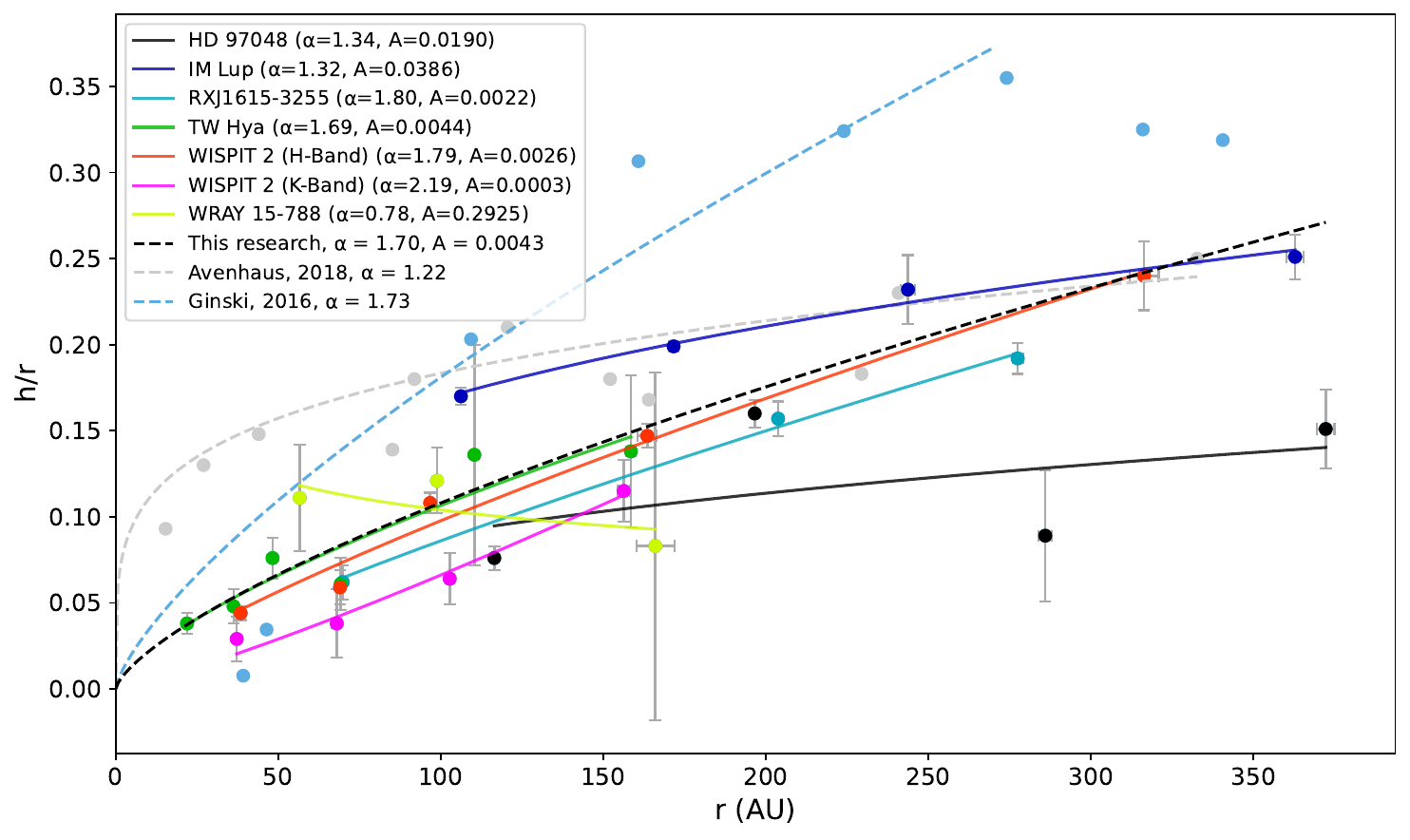}
        \caption{Aspect ratio ($h/r$) as a function of radius ($r$) for multi-ringed disks with 3 or more measurements only. Each disk is individually fitted with a power-law model ($h/r = A  \left(\frac{r}{1\mathrm{au}}\right)^{\alpha - 1}) $, with the resulting flaring index $\alpha$ indicated in the legend. Overlaid dashed lines represent the trend for extended disks (black), as well as comparisons to \citet{Ginski_2016} and \citet{Avenhaus2018}, in light blue and grey respectively.}
        \label{fig:wispitvsmulti.pdf}
    \end{figure*}

    From Fig.~\ref{fig:wispitvsmulti.pdf}, a range of different flaring profiles is observed for multi-ringed disks, which all show an increase in $h/r$ with separation (except for WRAY~15-788 \citep{Bohn2019}, which contains a significantly high error for the outermost ring). Excluding WRAY~15-788, IM~LUP \citep{Pinte2008} exhibits the lowest flaring index ($\alpha = 1.32$) but the highest overall flaring, with HD~97048 having the lowest flaring at large separations. V4046~Sgr \citep{Sissa2018} exhibits the steepest flaring index ($\alpha = 3.15$), although this value should be interpreted with caution due to the disk’s limited radial extent.

    RXJ~1615 \citep{de_Boer_2016} ($\alpha = 1.80$), TW~HYA \citep{vanBoekel2017} ($\alpha = 1.70$) and WISPIT~2 (in H-band; $\alpha = 1.79$) align with the flaring index found for extended disks ($\alpha = 1.72$) and \citet{Ginski_2016} for HD~97048 ($\alpha = 1.73$). 

    An interesting comparison case for multi-ringed systems is the newly identified disk around the young (approximately 5 Myr) solar-mass star WISPIT~2 (mass $= 1.08^{+0.06}_{-0.17}M_{\odot}$), recently resolved for the first time as part of the WISPIT (WIde Separation Planets In Time) survey. A detailed description of the system and its discovery is presented in \citet{VanCapelleveen2025}. Unlike the rest of the sample, for which geometric measurements were obtained in a single filter for each disk, WISPIT~2 was fitted in both H and K bands. This allows a more direct comparison of wavelength-dependent structure (see Appendix \ref{Wavelength Dependence Analysis} for the wavelength dependence analysis of the full sample), in which it is expected that longer wavelengths probe deeper scattering layers of dust within the disk. This makes WISPIT~2 a valuable case for detailed analysis, serving as a reference point for comparison against the general disk population.

    The geometric fits and the resulting properties measured can be found in \citet{VanCapelleveen2025}. The rings exhibit a mean inclination of $i = 43.99 \pm 0.87^\circ$ and a PA of $358.7 \pm 1.1^\circ$ in the H-band with $i = 45.86 \pm 1.27^\circ$, PA $= 359.3 \pm 2.3^\circ$ in K-band. Since the disk was only recently discovered and no prior ALMA measurements are available, the free ellipse fit was adopted. All structures were fitted in the H-band. In the K-band image, Ring 0 (the outermost ring) was too faint to be detected. 

    \begin{figure}[h!]
        \centering
        \includegraphics[width=0.9\linewidth]{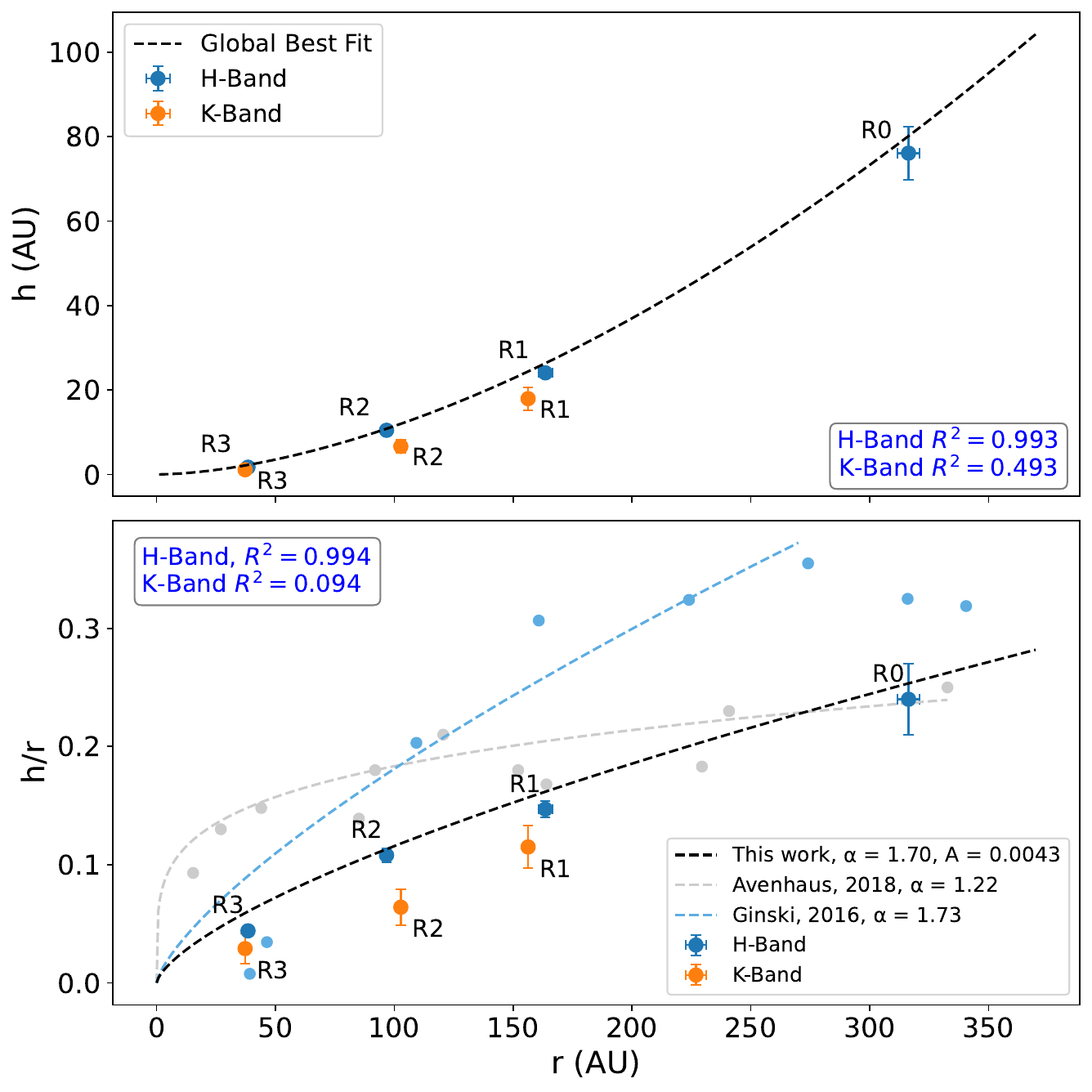}
        \caption{Vertical structure of the WISPIT~2 disk.
        \textbf{Top:} $h$ versus $r$ for H-band (blue) and K-band (orange). A power-law fit derived from the extended disk sample is overlaid (black).
        \textbf{Bottom:} $h/r$ versus $r$. The extended disk power law is again shown for comparison (black), \citet{Ginski_2016}, and \citet{Avenhaus2018}'s trends are also plotted in light blue and grey respectively. The $R^2$ values for both filter bands are shown to quantify the agreement with the extended disk trend for both height and aspect ratio. \textbf{Note:} In K-band, Ring 0 was not detected.}
        \label{fig:wispitgraphs.pdf}
    \end{figure}

    Notably, Ring 3 exhibits a significant asymmetry, with excess signal concentrated along the minor axis in the western direction and within the forward scattering section of the disk (investigated in detail in \citet{VanCapelleveen2025}). This localised brightness enhancement does not appear in the other rings. A similar signal excess is observed in R1 of RXJ~1615 (see Fig. \ref{fig:WISPvsRXJexcess}) and Ring 2 of HD~97048 \citep{Ginski_2016}, also both located along the disk's minor axis. These signal excesses are shown and further explored in Appendix~\ref{Signal Excesses}. Whether these asymmetries arise from intrinsic structural variations within the disk or result from the effects of polarimetric imaging remains uncertain. In the case of WISPIT~2, the discovery of WISPIT~2b adjacent to the excess suggests that the excess may be associated with the presence of a companion. Excesses pose clear challenges for edge detection if not properly identified and addressed during fitting. The SEEF algorithm can address this issue by allowing for the exclusion of specific azimuthal sections during analysis or by increasing the primary threshold level used. Since the algorithm is not fully automated and occurs iteratively, presenting the geometric detection output directly to the user for each pass, such aberrations are readily identifiable and were manually accounted for in the fitting process of the main sample. However, for Ring 3, Gaussian fitting failed to produce an azimuthally stable fit. To address this, the edge detection method was modified to identify the peak signal within the ring region (isolated via masking) rather than the outermost edge, effectively circumventing the issue.

    Fig.~\ref{fig:wispitgraphs.pdf} shows the heights and aspect ratios of each ring structure, plotted for both filters to assess their consistency/differences with the trend found for extended disks in Sect.~\ref{Global Results}. In the H-band, both the height and the aspect ratio show an exceptionally strong positive correlation to the extended disk trend, yielding coefficients of determination of $R_{\mathrm{h}}^2 = 0.993$ and $R_{\mathrm{h/r}}^2 = 0.975$, respectively. In contrast, the K-band shows a moderate correlation for height ($R_{\mathrm{h}}^2 = 0.501$), but the aspect ratio exhibits a weak negative correlation with radius ($R_{\mathrm{h/r}}^2 = -0.155$), suggesting that the K-band image is more affected by noise, as evident from the lower S/N ratio and the reduced number of detected structures. Limited radial coverage weakens the reliability of any inferred geometric trends in the absence of a measurement of Ring 0. In addition, opacity effects play a significant role: the longer wavelength of the K-band allows it to probe deeper into the disk, closer to the midplane, where the vertical height is intrinsically smaller. As seen in Fig.~\ref{fig:wispitgraphs.pdf}, both the measured heights and aspect ratios in K-band are systematically lower compared to the H-band, which traces higher scattering layers. This trend is clearly visible in the plots and aligns with expectations, supporting the interpretation that different wavelengths probe distinct vertical layers within the disk.

    The wavelength dependence of disk height and aspect ratio is a critical factor in interpreting the results of this study. To assess this effect, the analysis was repeated using only the H-band observations (see Appendix~\ref{Wavelength Dependence Analysis}), which showed no significant impact on the overall sample. However, as demonstrated by the results for WISPIT~2, this dependence must be considered in individual studies that include multi-wavelength data when comparing measurements.

    The evident flaring for both multi-ringed disks and extended disks (with most multi-ringed disks also being extended disks) suggests that the outer rings remain illuminated and observable, rather than being shadowed by the inner disk regions. If there is a population of disks that are not flared, the outer rings could be shadowed by the inner regions and therefore might not be observable in scattered light with current limitations. This suggests that the presence of multiple visible rings provides indirect evidence for a flared geometry across these disks.

\subsection{System Properties affecting the Vertical Height Structure}
\label{System Properties affecting the Vertical Height Structure}

    To explore whether disk flaring correlates with broader system properties, the aspect ratios derived for the full sample (see Appendix~\ref{Appendix Heights} for all measured values) were compared against dust mass, stellar age, and stellar mass. These stellar and disk properties were computed by Garufi et al. (in prep.), following the methodology described in \citet{Garufi2024}. The spectral energy distribution and effective temperature was collected for each source through Vizier. Following this, the stellar luminosity was calculated from a de-reddened Pheonix model of the stellar photosphere \citep{Hauschildt1999}. The stellar age and mass were then constrained from isochrones \citep{Siess2000, Bressnan2012, Baraffe2015}. The dust mass was directly from the observed millimetre photometry of each source with standard assumptions on optically thin emission and dust opacity \citep{Beckwith1990}. Not all disks within this work's dataset are represented in Garufi et al. (in prep.)'s analysis. Stellar masses and ages were available for 61 out of the 92 unique disks, while dust masses were reported for 46 disks.
    
\begin{figure}[h!]
        \centering
        \includegraphics[width=1\linewidth]{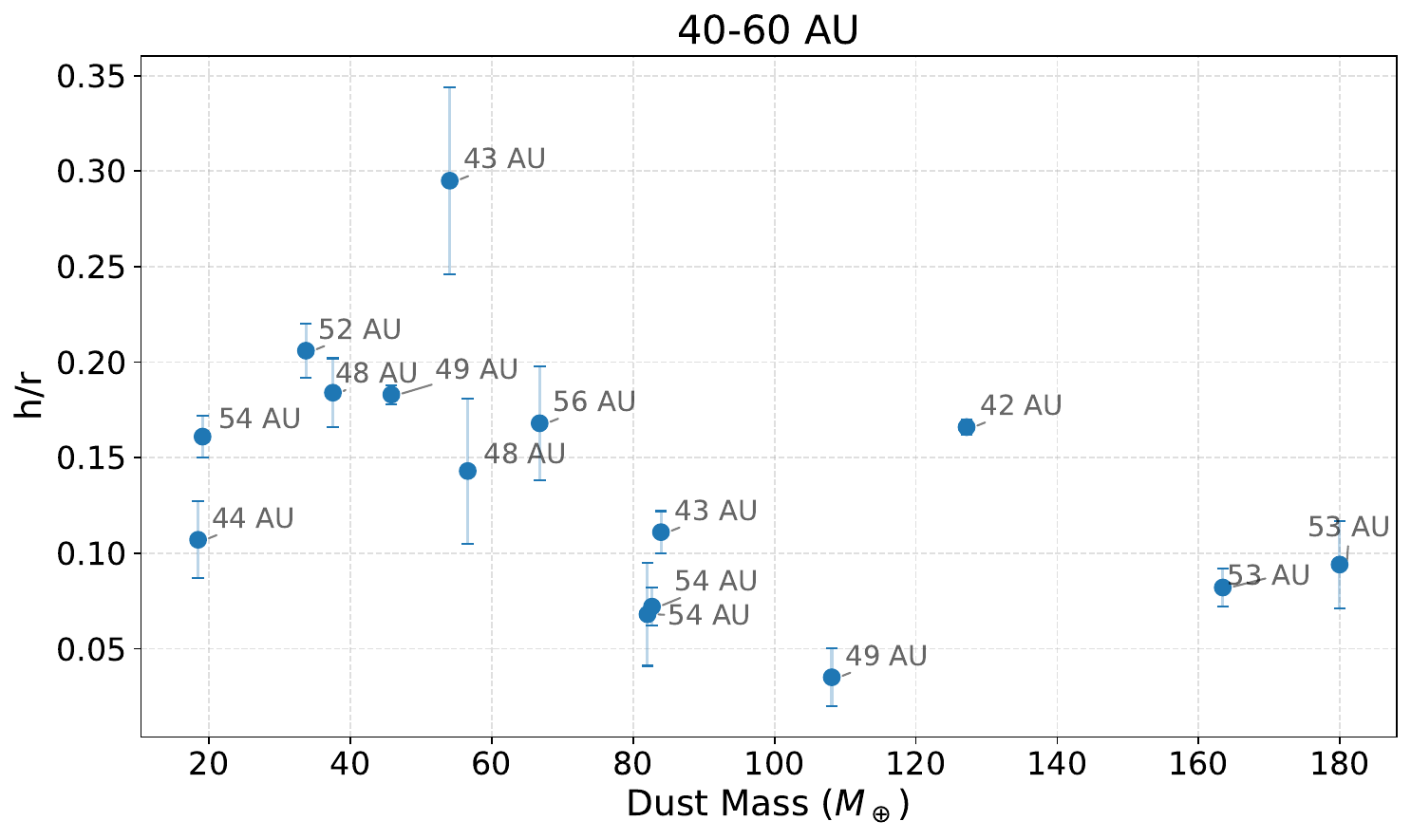}
        \caption{Aspect ratio ($h/r$) versus disk dust mass (in Earth masses) for all measurements within a separation bracket of 40 - 60~au. No significant correlation is observed, indicating that within this radial range (and found for all other brackets), the vertical structure does not show a clear dependence on the total dust mass of the system.}
        \label{fig:dustmass}
    \end{figure}

    Ideally, all measurements should be taken at similar radial distances since $h/r$ increases with radius, comparing measurements at different radii can introduce scatter unrelated to the investigated property. The plot shown in Fig.~\ref{fig:dustmass} explores the potential relationship between aspect ratio and dust mass, using only measurements within a radius range between 40 and 60~au, as this bracket contained the most measurements (15 unique disks out of 46 in the total sample). However, for all brackets with a reasonable number of data points ($n > 5$), no correlation has been found.

    Within this dataset, there are insufficient data points at common radial distances (ideally within 5~au of each other) to construct a meaningful comparison. Another factor to consider is the high uncertainty in the calculation of the dust mass. A gas-to-dust mass is assumed to be the same as the interstellar medium (ISM) with a ratio of 100 \citep{Bohlin1978}. However, if the gas/dust ratio varies between disks, the aspect ratio/height profile is expected to vary also. \citet{Williams2014} found a wide range of inferred gas/dust ratios for nine class II disks with a mean ratio value of 16, with all ratios below the assumed ISM ratio. Further, the dust mass calculation assumes the dust is optically thin. In reality, a significant fraction of the disk can be optically thick at millimetre wavelengths \citep{Garufi2025}. This particularly affects compact disks as a larger fraction of the material along the line of sight is concentrated in a small area, leading to an underestimation of dust mass that is stronger for smaller disks. Thus, the results presented in Fig.~\ref{fig:dustmass} may contain a trend that is concealed by the effect of the opacity.
    
\begin{figure}[h!]
        \centering
        \includegraphics[width=1\linewidth]{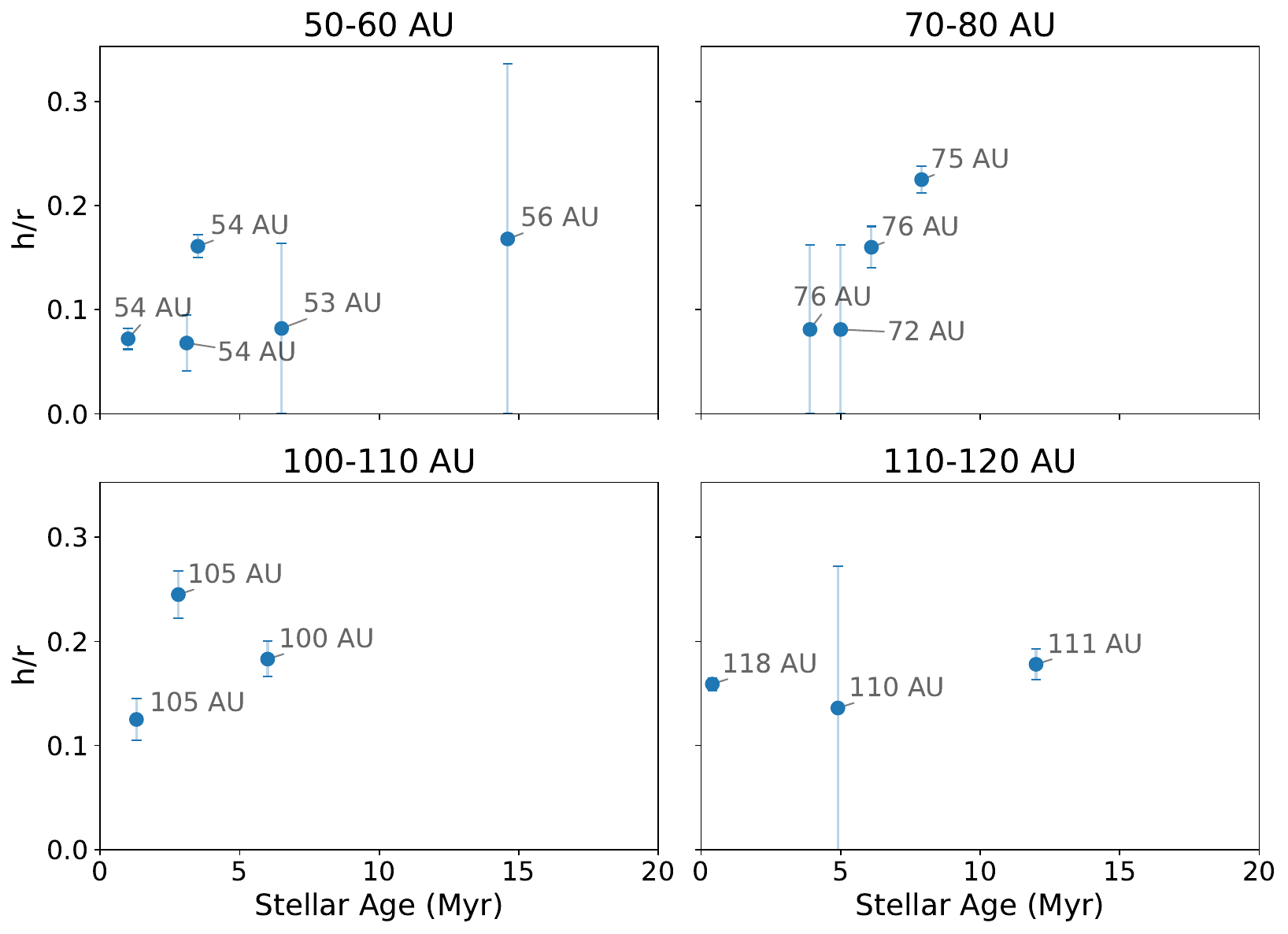}
        \caption{Aspect ratio ($h/r$) plotted against stellar age (Myr) for four separation brackets, to examine how disk vertical structure evolves with age.}
        \label{fig:age_hr}
    \end{figure}

    Fig.~\ref{fig:age_hr} shows the aspect ratio vs. the stellar age for four different radial distance brackets of 10~au. Stellar age is assumed to represent the time since disk formation, with the underlying premise that the circumstellar disk forms simultaneously with its host star. The aspect ratio of protoplanetary disks is expected to decrease with stellar age as the disk cools and its dust and gas dissipate, leading to reduced pressure support and less flared geometries. Fig.~\ref{fig:age_hr} reveals no significant correlation found between stellar age and aspect ratio for any single bracket. However, while understanding that this sample represents small number statistics and the correlation is weak, it is noted that the lowest aspect ratios are found for the youngest systems and the older systems typically have larger aspect ratios. A plausible explanation for this is the effect of "self-shadowing".

    Self-shadowing is a geometric effect in circumstellar disks where the inner regions block starlight from reaching the outer parts of the disk. This typically occurs when the disk has a "puffed-up" or vertically extended inner rim, often caused by direct irradiation, which casts a shadow over the outer disk \citep{Natta2001,Dullemond_2001}. As a result, the shadowed region receives lower stellar flux, leading to lower temperatures and flatter structures. Self-shadowing can significantly alter the observed brightness, thermal emission, and vertical structure of a disk, and is a key factor in explaining non-uniform illumination patterns seen in scattered light images. Assuming as the disk evolves, through planet formation carving gaps or other effects such as photoevaporation or accretion, the disk begins to carve an inner cavity (i.e. a transition disk), the disk would become less self-shadowed and the temperature of the outer regions of the disk would increase leading to a greater aspect ratio for older systems. \citet{garufi2018} found that younger disks ($\le 5$ Myr) are systematically fainter than older disks in scattered light and \citet{Garufi2022} did not find a significant trend but noted that young sources ($ \le 2$ Myr) were averagely fainter than older sources. Should this increase in brightness in older systems be caused by the reduction of self-shadowing, this would lead to greater aspect ratios in older disks in agreement with the weak correlation between aspect ratio and stellar age found here, despite the relationship of a decreasing dust mass with age \citep{garufi2018}.

    Considering heights were measured at the most outward radial extent with the edge detection method and this method was applied to all continuous disks not containing gaps (i.e. the majority of disks with r $\lessapprox 150$~au), if self-shadowing plays a critical role in the aspect ratio scatter observed in Sect. \ref{Global Results}, then this could explain the large scatter found at these radial extents. As \citet{Garufi2024} note, the inner rim is expected to lie within 0.02–2~au of the star; therefore, the measured feature cannot correspond to the inner rim. However, if a moderate fraction of these disks are partially self-shadowed, this effect could explain the wide spread in measured aspect ratios at radii $\lessapprox$~150 au. Extended disks do not exhibit this scatter, suggesting that if self-shadowing is indeed responsible, it is not present in these systems. This is consistent with the interpretation that such disks have cleared their inner cavities (i.e. Group I or transition disks), reducing shadowing and allowing for more uniform flaring, in agreement with theoretical expectations \citep{Natta2001}. These systems may have evolved beyond the self-shadowed phase (i.e. cleared the inner region), or never experienced it to a significant degree. Conversely, more compact disks may be preferentially affected by self-shadowing, suppressing/increasing the aspect ratio significantly due to their relatively smaller intrinsic size and thereby introducing the observed scatter at small separations ($r < 50$~au). Further investigation into the role of self-shadowing would confirm if this is indeed a factor within the observed scatter in vertical height between the disks in this dataset.

    The same analysis was repeated for stellar mass which is expected to correlate with the vertical height (see Appendix \ref{StellarMassCalc}), assuming that the classical MLR does not hold for pre main-sequence stars. However, similar to the stellar age, no correlation was found for aspect ratio vs. stellar mass.

    For all of these previous analyses, a key limitation is the need to measure aspect ratios at a consistent radius across systems, which drastically reduces the sample size for each analysis, leading to small-number statistics. Additionally, if varying disk morphologies exhibit different aspect ratios, this would complicate any simple relationship between the dust mass, stellar age or stellar mass and the disk geometry.

\subsection{Planets Opening Gaps}
\label{Planets Opening Gaps}

    The formation of gaps in protoplanetary disks by embedded planets is a well-established outcome of planet-disk interactions, driven by the exchange of angular momentum between the planet and the disk gas \citep{lin1979L,Goldreich1980,Lin1993}. The relation between planet mass and gap width has been extensively explored through hydrodynamic simulations \citep{Kanagawa2016,Dong__2017,Zhang2018}. Given that both gap width and aspect ratio are measured through ellipse fitting and are key parameters in planet-disk interaction models, the results of the SEEF algorithm provide a natural foundation for exploring the theoretical masses of planets responsible for opening these gaps.
    
    In this work, the empirical model developed by \citet{Kanagawa2016} is applied to estimate planet masses from observed gap widths, assuming the gap is opened by a single planet located at the gap centre:
    
    \begin{equation}
        \frac{M_\mathrm{p}}{M_*} = 2.1 \times 10^{-3} \left(\frac{\Delta_{\mathrm{gap}}}{R_\mathrm{p}}\right)^2\left(\frac{h_\mathrm{p}}{0.05R_p}\right)^\frac{3}{2}\left(\frac{\alpha_{\mathrm{visc}}}{10^{-3}}\right)^{\frac{1}{2}},
    \end{equation}
    
    where $M_\mathrm{p}$, $R_\mathrm{p}$, and $h_\mathrm{p}/R_\mathrm{p}$ are the planet mass, orbital radius, and local disk aspect ratio of the gas pressure scale height, respectively; $M_*$ is the stellar mass; and $\alpha_{\mathrm{visc}}$ is the disk viscosity parameter \citep{shakura1973}. Following \citet{Ginski_2016, ginski2025}, the gas pressure scale height is estimated from the dust scattering surface height by dividing by a factor of 4, under the assumption that gas and dust are well mixed \citep{Chiang2001}. We note that the conversion between the scattering-surface height and the gas pressure scale height (here assumed to be a factor of four) is uncertain and depends on the dust optical depth and vertical dust distribution, introducing a systematic uncertainty in the inferred gas scale height, and consequently in the estimated planet masses. For each gap, the table lists the gap width ($\Delta_{\mathrm{gap}}$), the corresponding planet’s semi-major axis ($R_\mathrm{p}$), and the local disk aspect ratio, derived from observations. The gap width was calculated as $R_0 - R_1$, where $R_0$ and $R_1$ are the radii of the adjacent rings, measured along the major-axis and $R_0$ is the most outward ring of the pair. This definition formally describes the gas gap; throughout this work, we assume that the dust is sufficiently well coupled to the gas that the measured dust ring locations trace the underlying gas structure, consistent with the assumptions adopted in \citet{Kanagawa2016}. This assumption neglects the fact that gaps seen in scattered light may also be shaped by shadowing and by the geometry of the disk scattering surface. The planet is assumed to orbit at the midpoint of the gap. The aspect ratio was calculated using the scale parameter \(A\) and flaring index \(\alpha\) derived from the power-law fits for individual systems (see Fig.~\ref{fig:wispitvsmulti.pdf}). For systems with fewer than three height measurements, the extended disk trend was adopted instead (see Sect.~\ref{Global Results}). For WRAY~15-788 with three measurements, which had a large uncertainty for the outermost ring, the extended disk trend values were used. These values were then applied using the power-law relation:
    
    \begin{equation}
        \frac{h_\mathrm{p}}{R_\mathrm{p}} = A\left(\frac{R_\mathrm{p}}{1\mathrm{au}}\right)^{\alpha-1}.
    \end{equation}

    \begin{figure}[h!]
        \centering
        \includegraphics[width=1\linewidth]{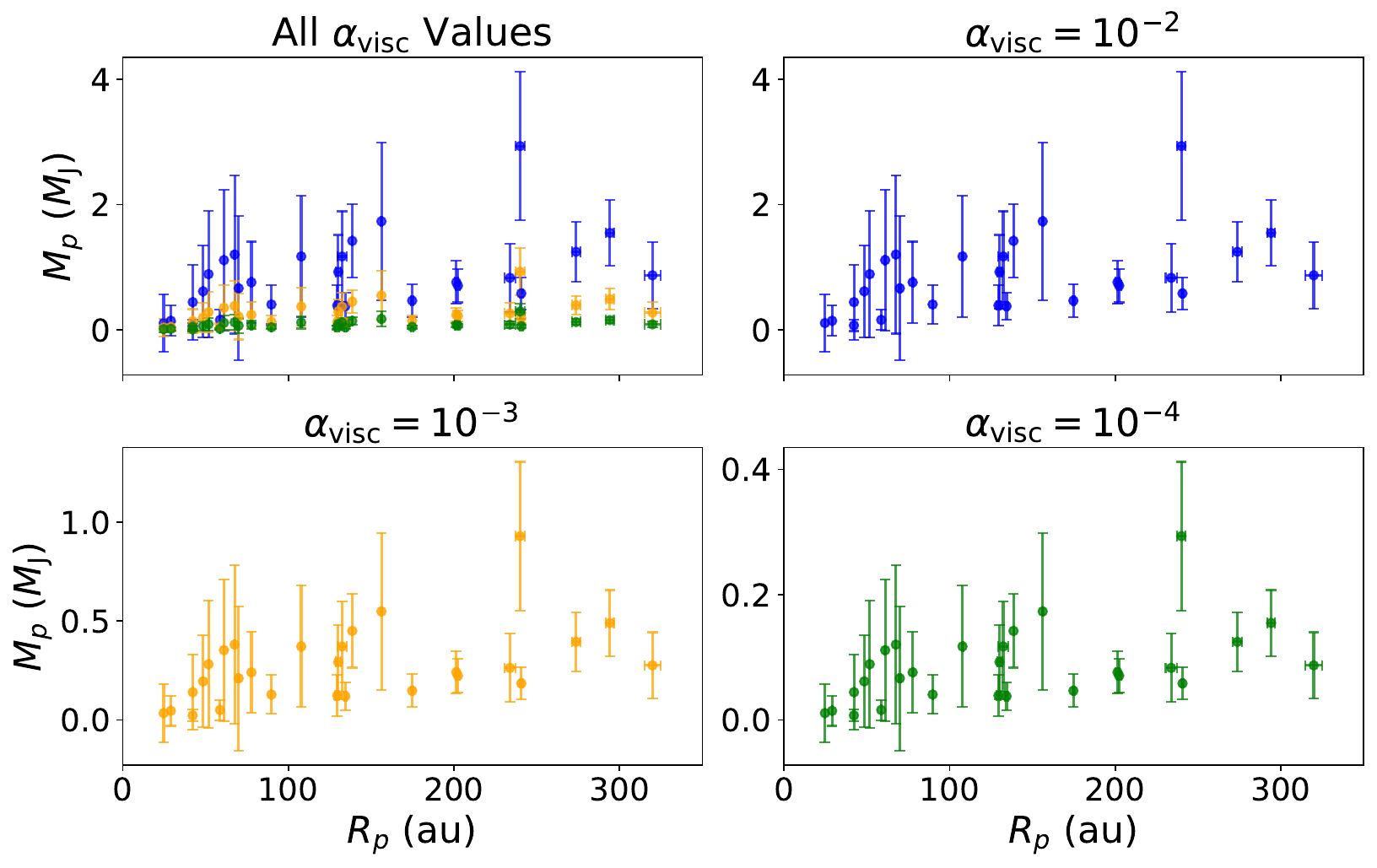}
        \includegraphics[width=1\linewidth]{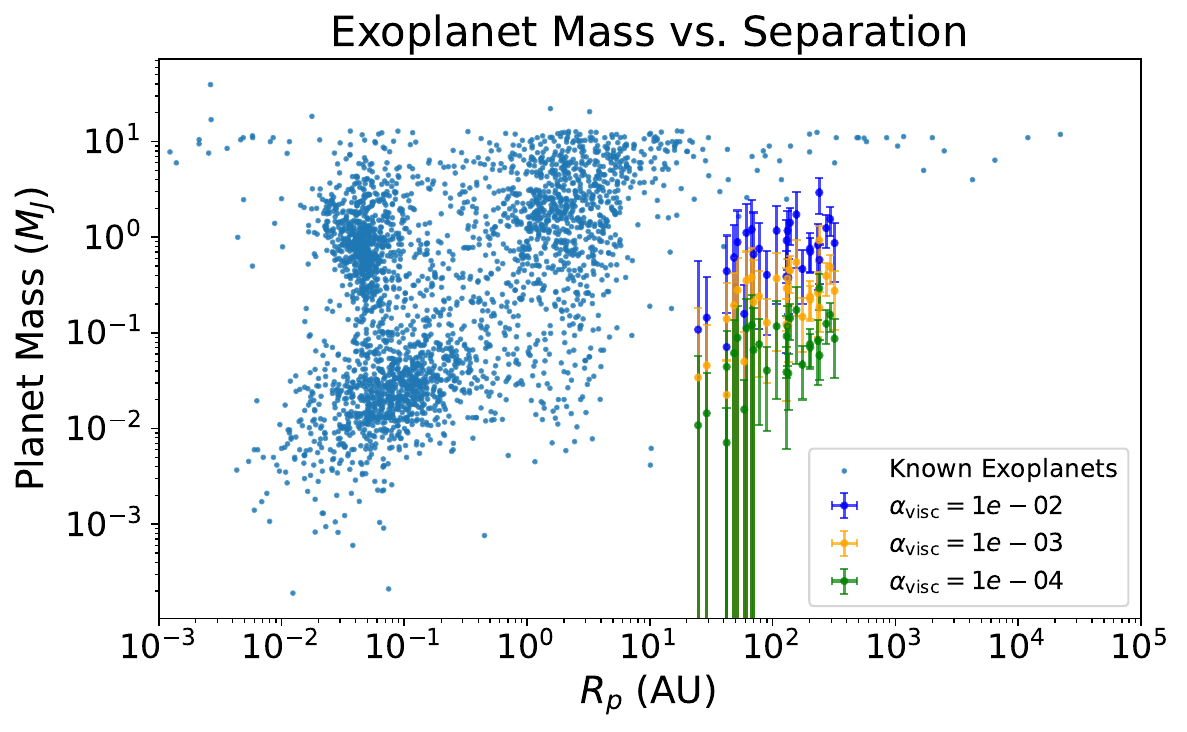}
        \caption{\textbf{Top:} Planet mass ($M_\mathrm{J}$) versus separation of the inferred planet ($R_\mathrm{p}$, au) for three different values of the disk viscosity parameter $\alpha_{\mathrm{visc}}$. The top-left panel shows all $\alpha_{\mathrm{visc}}$ values overlaid together. The remaining panels show individual plots for $\alpha_{\mathrm{visc}}$ = $10^{-2}$, $10^{-3}$, and $10^{-4}$, respectively. Error bars represent uncertainties in both the inferred planet mass and the measured gap radius. The plot highlights the dependence of inferred planet mass on $\alpha_{\mathrm{visc}}$, with lower viscosity requiring lower planet masses to carve equivalent gaps.   
        \textbf{Bottom:} Distribution of detected exoplanets plotted as mass ($M_J$) vs. separation (au). Blue points represent confirmed exoplanets from the \texttt{exoplanet.eu} database, where true masses are used when available and minimum masses ($M \times\sin i$) are used otherwise. Other coloured points correspond to the planet masses inferred for different viscosities in this work.}
        \label{fig:alphachanged}
    \end{figure}
    
    Planet masses are calculated under the assumption that the gaps are opened by single planets and are reported for three values of the disk viscosity parameter $\alpha_{\mathrm{visc}} = 10^{-2}$, $10^{-3}$, and $10^{-4}$ to reflect different plausible disk conditions. Uncertainties were estimated as follows: the gap width uncertainty reflects the propagated uncertainties in the positions of the adjacent rings; a 10\% uncertainty was assigned to the stellar mass; and a fixed uncertainty of 0.05 was adopted for the aspect ratio, consistent with the scatter observed in the extended disk population (see Fig.~\ref{fig:extended3}). These estimates provide a comparative view of how inferred planet masses vary with assumed disk viscosity. Fig.~\ref{fig:alphachanged} shows the inferred planet masses for the sample of multi-ringed disks for the three values of disk viscosity.

    The bottom panel of Fig.~\ref{fig:alphachanged} presents the inferred planet masses and separations derived, overlaid on the broader population of detected exoplanets for reference. Among the models explored, the high viscosity $\alpha_{\mathrm{visc}} = 10^{-2}$ case aligns most closely with the current exoplanet detections in the mass-separation plane yet still below current detection limits. However, this apparent agreement should be interpreted with caution, as planet migration may shift planets inward from their original formation locations, meaning that the observed disk gaps may not correspond directly to current orbital positions \citep{Goldreich1980,Fernandez1984}. In the low-viscosity cases ($\alpha_{\mathrm{visc}} = 10^{-3}$ and $10^{-4}$), the inferred planets occupy a largely unexplored region of parameter space, potentially due to significant inward migration after formation. It is noted that this region of the mass-separation plane aligns with the population of debris-inferred exoplanets using sculpting and stirring arguments \citep{Pearce2022}.

    For disk viscosity parameters between $10^{-2}$ and $10^{-4}$, the range of inferred planet is between approximately $2M_J$ and $0.2M_J$ for larger gaps or $0.02M_J$ and $0.2M_J$ for smaller gaps. Observationally, current detection limits within disk gaps are primarily constrained by achievable contrasts. \citet{ginski2025} derived a limit of approximately $5~M_{J}$ for the gap in 2MASS~J1612, while \citet{TTorres2021} reported a range of detection limits from $9~M_{J}$ at close separations (approximately 10~au) down to $4~M_{\rm Jup}$ at wider separations (approximately 100~au), highlighting how sensitivity improves with increasing orbital radius. Considering the detection limits, then all the inferred planets would not be directly detectable by current limitations. However, the large uncertainties in the inferred planet masses (see Table \ref{GapTable}), often exceeding the nominal values, arise primarily from the $\pm 0.05$ uncertainty in the aspect ratio, so these results should be interpreted more as an investigation into the dependence of the inferred planet masses on the viscosity parameter than an estimate of the true planet masses.

\section{Summary and Conclusions}
\label{Conclusions}

    This study analyses the vertical height structure for a sample of 92 class-II systems containing disks expanding on the previous literature for vertical height measurement \citep{de_Boer_2016,Ginski_2016,Avenhaus2018,Rich2021,Ginski2024}. It also presents a methodology, expanding on the methods introduced by \citet{Ginski2024}, for robustly measuring the vertical heights and aspect ratios for differing morphologies of protoplanetary disks. The vertical heights, aspect ratios, separations, inclinations and position angles of the structures with the 92 disks are found within Table \ref{AllTable} in Appendix \ref{Appendix Heights}.

    For the full sample, including the investigation into removing spiral, face-on and inclined disks and further and limiting the dataset to H-band filter images only (see Appendix \ref{Wavelength Dependence Analysis}), it is found that the vertical height structure cannot be confidently described by a single power law. This is evidenced by the scatter and moderate-to-low correlation observed in the $h/r$ vs. $r$ distributions. The impact of observational filter used on the inferred flaring is found to be significant as seen in the case of WISPIT~2 and must be considered when comparing vertical height measurements made using different filters. However, in this work, where the majority of images are in H-band, the choice of filter does not significantly influence the trends presented. The low correlations suggest that either (1.) different disk morphologies intrinsically follow distinct vertical structures, or (2.) an additional, unidentified factor contributes to the scatter across the population. Notably, the categorical breakdowns and system properties (Sect. \ref{System Properties affecting the Vertical Height Structure}) explored in this work revealed no significant trends within any underlying factor or subgroup, except for the extended disks, which remain the only category consistently following a well-defined flaring power law. 

    Extended disks ($r_{outer} \ge 150$au) are found to be well-described by a single power law:

    \begin{equation}
        h = 0.0043\left(\frac{r}{1\mathrm{au}}\right)^{1.70},
    \end{equation}
    
    suggesting that some underlying physical property of these systems is resilient to the factors driving the observed spread in flaring among other disk categories. This consistency implies that extended disks may occupy a distinct evolutionary or structural regime in which processes such as dust settling, turbulence, or self-shadowing are either stabilised or play a diminished role in shaping the vertical structure.

    While no statistically significant correlation was found between aspect ratio and disk properties such as stellar mass, age, or dust mass, a trend is noted in which the youngest systems exhibit the lowest aspect ratios, with older systems showing higher values. If self-shadowing plays a dominant role in shaping vertical height and disk flaring, this trend would be consistent with older disks clearing their inner region, i.e., Group I disks. However, the investigation into these dependencies is strongly limited by small-number statistics. Disk self-shadowing may contribute to the observed scatter in vertical height, independent of stellar mass, age, disk category, or dust mass.

    Using the vertical height trends, the masses of gap-opening planets are estimated. The inferred planet masses span approximately $0.2$–$2\,M_{\mathrm J}$ for larger gaps and $0.02$–$0.2\,M_{\mathrm J}$ for smaller gaps, depending on the assumed disk viscosity parameter ($\alpha_{\mathrm{visc}}$). These planets remain below current direct-detection limits. Given the large uncertainties in these estimates, the derived values should therefore be regarded as tentative. With reference to the detected exoplanet population (Fig.~\ref{fig:alphachanged}) and assuming the traced disk gaps are pre-dominantly opened by planets forming gaps, our analysis suggests two potential scenarios. If the planets remain at their current location in the mass-separation plane, then the disk structures trace an entirely new and lower-mass population of wide-separation planets compared to those previously detected. Alternatively, it may be possible that these planets are still undergoing significant inward migration within the disk, in which case these may be pre-cursors to the lower separation planet population detected with indirect methods.

\begin{acknowledgements}
    We would like to thank the anonymous reviewer for their careful reading of the manuscript and for the many constructive comments that helped improve the quality of this work.
    This work is based on observations collected with the SPHERE instrument on the Very Large Telescope at the European Southern Observatory under multiple ESO programmes, retrieved from the ESO Science Archive Facility. The research has made use of the following software and Python libraries: NumPy \citep{Harris2020}, SciPy \citep{2020SciPy-NMeth}, and Astropy \citep{astropy:2013,astropy:2018,astropy:2022}. The author acknowledges the support and supervision provided by the School of Natural Sciences and the Centre for Astronomy at the University of Galway. This paper is based on the author’s Master’s thesis carried out at the University of Galway.
\end{acknowledgements}

\bibliographystyle{aa} 
\bibliography{refs} 

\clearpage

    \begin{appendix}
    \onecolumn
    \section{Table of Results}
    \label{Appendix Heights}
    
    Table \ref{AllTable} details the height (au), inclination (degrees), separation (au), aspect ratio ($h/r$), PA (degrees) and whether the ellipse fit had prior estimates of the inclination and PA for the constrained ellipse fit (Constr?) for all measured disks. The region of the ellipse fit is noted to differentiate between substructures of the same disk, where inner (2) and (3) refer to inner rings at a greater radial extent than the most inner structure which is simply referred to as inner. (*) The PA shown is measured from the +y axis to the first instance of the major axis counter-clockwise, as the front scattering side could not be determined.

    \begin{longtable}{l l c c c c c c c}
    \caption{Geometrical parameters of the disks.}\label{AllTable} \\
    \toprule
    Disk & Region & Height (au) & Incl (deg) & Separation (au) & $h/r$ & PA (deg) & Constr? \\
    \midrule
    \endfirsthead
    \caption{Continued.} \\
    \toprule
    Disk & Region & Height (au) & Incl (deg) & Separation (au) & $h/r$ & PA (deg) & Constr? \\
    \midrule
    \endhead
    \midrule
    \multicolumn{8}{r}{{\em Continued on next page}} \\
    \midrule
    \endfoot
    \bottomrule
    \endlastfoot
    J13342461-6517473 & Inner & $2.36 \pm 0.52$ & $43.68 \pm 1.94$ & $28.99 \pm 0.80$ & $0.08 \pm 0.02$ & $114.14 \pm 3.59$ & No \\
    J16042165-2130284 & Inner & $9.25 \pm 0.76$ & $15.10 \pm 0.01$ & $78.24 \pm 0.16$ & $0.12 \pm 0.01$ & $352.10 \pm 3.22$ & Yes \\
     & Outer & $8.64 \pm 3.68$ & $10.57 \pm 0.04$ & $113.16 \pm 0.57$ & $0.08 \pm 0.03$ & $343.63 \pm 23.23$ & Yes \\
    J16082324-1930009 & Outer & $3.28 \pm 0.84$ & $70.16 \pm 1.23$ & $61.51 \pm 1.76$ & $0.05 \pm 0.01$ & $303.88 \pm 0.88$ & No \\
    J16083070-3828268 & Outer & $18.39 \pm 0.54$ & $73.93 \pm 0.31$ & $145.57 \pm 1.65$ & $0.13 \pm 0.00$ & $287.42 \pm 0.27$ & No \\
    J16090075-1908526 & Inner & $1.36 \pm 0.44$ & $52.46 \pm 2.01$ & $39.29 \pm 1.25$ & $0.04 \pm 0.01$ & $144.13 \pm 3.97$ & No \\
    J16120668-3010270 & Inner & $2.93 \pm 1.54$ & $12.31 \pm 3.04$ & $29.97 \pm 0.23$ & $0.10 \pm 0.05$ & $227.25 \pm 33.18$ & Yes \\
     & Middle & $16.93 \pm 0.96$ & $35.42 \pm 0.92$ & $75.09 \pm 0.39$ & $0.23 \pm 0.01$ & $221.04 \pm 1.58$ & Yes \\
    J18521730-3700119 & Inner & $1.25 \pm 0.21$ & $38.76 \pm 0.63$ & $24.50 \pm 0.11$ & $0.05 \pm 0.01$ & $299.08 \pm 0.71$ & Yes \\
     & Outer & $3.37 \pm 0.79$ & $34.74 \pm 1.03$ & $54.16 \pm 0.36$ & $0.06 \pm 0.01$ & $300.36 \pm 1.60$ & Yes \\
    J19005804-3645048 & Outer & $0.44 \pm 0.07$ & $65.61 \pm 0.75$ & $36.72 \pm 0.55$ & $0.01 \pm 0.00$ & $100.67 \pm 88.75$ (*) & No \\
    AK     Sco & Outer & $3.24 \pm 1.91$ & $63.53 \pm 5.27$ & $29.45 \pm 1.01$ & $0.11 \pm 0.07$ & $231.36 \pm 1.08$ & Yes \\
    AS     209 & Inner & $6.60 \pm 1.61$ & $36.81 \pm 4.44$ & $35.67 \pm 0.78$ & $0.18 \pm 0.04$ & $105.67 \pm 35.93$ & Yes \\
    AT     Pyx & Inner & $21.41 \pm 1.90$ & $36.39 \pm 1.22$ & $179.97 \pm 0.73$ & $0.12 \pm 0.01$ & $129.37 \pm 0.78$ (*) & No \\
    BD-06          1178 & Outer & $17.51 \pm 0.79$ & $61.07 \pm 0.89$ & $79.12 \pm 1.15$ & $0.22 \pm 0.01$ & $30.68 \pm 0.93$ & No \\
    BD-08          1115 & Inner & $22.49 \pm 1.22$ & $63.54 \pm 1.78$ & $63.15 \pm 0.76$ & $0.36 \pm 0.02$ & $247.91 \pm 0.34$ & Yes \\
    BN     CrA & Inner & $5.83 \pm 1.18$ & $66.60 \pm 0.68$ & $71.52 \pm 1.51$ & $0.08 \pm 0.02$ & $298.09 \pm 0.55$ & No \\
    CD-38          4381 & Outer & $11.96 \pm 2.16$ & $28.11 \pm 1.85$ & $86.94 \pm 1.09$ & $0.14 \pm 0.03$ & $66.38 \pm 50.00$ (*) & No \\
    CM     Cha & Outer & $19.90 \pm 7.05$ & $18.83 \pm 3.41$ & $72.28 \pm 0.92$ & $0.28 \pm 0.10$ & $123.45 \pm 26.84$ & Yes \\
    CQ     Tau & Inner & $4.78 \pm 0.48$ & $50.15 \pm 0.97$ & $42.93 \pm 0.32$ & $0.11 \pm 0.01$ & $56.06 \pm 0.51$ & Yes \\
    CS     Cha & Inner & $3.64 \pm 1.45$ & $28.81 \pm 1.35$ & $53.62 \pm 0.78$ & $0.07 \pm 0.03$ & $74.61 \pm 7.88$ & Yes \\
    CT     Cha & Outer & $6.73 \pm 1.16$ & $48.77 \pm 3.13$ & $41.77 \pm 1.03$ & $0.16 \pm 0.03$ & $58.78 \pm 1.96$ & Yes \\
    CV     Cha & Inner & $8.82 \pm 0.83$ & $35.93 \pm 1.88$ & $48.00 \pm 0.53$ & $0.18 \pm 0.02$ & $312.78 \pm 3.43$ & Yes \\
    DG     CrA & Outer & $7.48 \pm 1.04$ & $36.94 \pm 0.83$ & $170.03 \pm 0.75$ & $0.04 \pm 0.01$ & $279.08 \pm 28.36$ & No \\
    DG     Tau & Outer & $31.64 \pm 10.62$ & $25.59 \pm 4.87$ & $154.41 \pm 3.52$ & $0.20 \pm 0.07$ & $132.97 \pm 4.38$ & Yes \\
    DH     Tau & Inner & $1.72 \pm 1.19$ & $25.14 \pm 3.54$ & $35.59 \pm 0.76$ & $0.05 \pm 0.03$ & $131.45 \pm 14.65$ (*) & Yes \\
    DL     Tau & Outer & $1.70 \pm 0.72$ & $43.07 \pm 7.88$ & $48.72 \pm 2.09$ & $0.04 \pm 0.01$ & $52.14 \pm 6.57$ & Yes \\
    DM     Tau & Inner & $6.93 \pm 1.84$ & $33.59 \pm 1.95$ & $48.38 \pm 0.60$ & $0.14 \pm 0.04$ & $155.99 \pm 4.83$ & Yes \\
    DO     Tau & Outer & $33.25 \pm 1.13$ & $36.17 \pm 1.09$ & $58.85 \pm 0.59$ & $0.56 \pm 0.02$ & $156.19 \pm 16.36$ & No \\
    DoAr     44 & Outer & $12.53 \pm 2.08$ & $49.49 \pm 3.11$ & $42.52 \pm 1.00$ & $0.29 \pm 0.05$ & $57.47 \pm 1.76$ (*) & Yes \\
    DQ     Tau & Inner & $1.22 \pm 0.20$ & $45.30 \pm 2.47$ & $32.73 \pm 0.55$ & $0.04 \pm 0.01$ & $38.33 \pm 1.99$ & Yes \\
    DW     Aur & Outer & $4.15 \pm 0.50$ & $62.29 \pm 1.75$ & $49.91 \pm 1.02$ & $0.08 \pm 0.01$ & $307.50 \pm 1.79$ & No \\
    EM     AS          218 & Inner & $2.50 \pm 0.25$ & $24.41 \pm 0.97$ & $34.29 \pm 0.21$ & $0.07 \pm 0.01$ & $66.93 \pm 3.00$ & No \\
     & Outer & $5.89 \pm 9.23$ & $31.59 \pm 1.40$ & $76.51 \pm 0.63$ & $0.08 \pm 0.12$ & $58.95 \pm 3.56$ & No \\
    EM LkCa     15 & Inner & $5.72 \pm 0.42$ & $40.82 \pm 1.79$ & $38.08 \pm 0.31$ & $0.15 \pm 0.01$ & $66.67 \pm 0.71$ & Yes \\
    EM     LkHA          330 & Outer & $23.30 \pm 2.20$ & $41.05 \pm 0.57$ & $145.89 \pm 0.80$ & $0.16 \pm 0.01$ & $231.72 \pm 1.36$ & Yes \\
    EM     LkHA          345 & Outer & $8.35 \pm 0.96$ & $29.30 \pm 1.96$ & $39.72 \pm 0.46$ & $0.21 \pm 0.02$ & $160.61 \pm 5.14$ & No \\
    EM     SR               21A & Outer & $18.16 \pm 1.90$ & $19.10 \pm 1.20$ & $75.86 \pm 0.38$ & $0.24 \pm 0.03$ & $30.99 \pm 5.95$ (*) & Yes \\
    EM          SR4 & Outer & $12.85 \pm 2.72$ & $27.22 \pm 4.53$ & $48.81 \pm 0.58$ & $0.26 \pm 0.06$ & $20.21 \pm 9.44$ & No \\
    Gaia     DR2     58548- & Inner & $6.99 \pm 0.64$ & $34.97 \pm 1.11$ & $27.05 \pm 0.24$ & $0.26 \pm 0.02$ & $199.41 \pm 2.92$ & No \\
    97321965963264 \\
    GG Tau A & Outer & $41.42 \pm 1.88$ & $39.24 \pm 0.59$ & $207.79 \pm 0.93$ & $0.20 \pm 0.01$ & $104.36 \pm 0.78$ & Yes \\
    GM     Aur & Outer & $13.06 \pm 2.06$ & $60.87 \pm 3.47$ & $104.78 \pm 1.25$ & $0.12 \pm 0.02$ & $57.85 \pm 0.81$ & Yes \\
    Haro     1-1 & Outer & $6.12 \pm 1.79$ & $50.08 \pm 0.79$ & $75.87 \pm 0.62$ & $0.08 \pm 0.02$ & $225.14 \pm 1.56$ & No \\
    Haro     5-34 & Outer & $2.99 \pm 1.41$ & $11.34 \pm 4.72$ & $128.50 \pm 1.30$ & $0.02 \pm 0.01$ & $27.65 \pm 40.64$ & No \\
    HD     100453 & Outer & $7.01 \pm 0.17$ & $31.58 \pm 0.38$ & $42.18 \pm 0.12$ & $0.17 \pm 0.00$ & $2.06 \pm 0.91$ & No \\
    HD 100546 & Inner & $18.35 \pm 1.69$ & $49.24 \pm 2.26$ & $100.41 \pm 1.09$ & $0.18 \pm 0.02$ & $324.01 \pm 1.33$ & Yes \\
    HD 143006 & Outer & $3.67 \pm 1.77$ & $16.25 \pm 7.67$ & $85.82 \pm 2.92$ & $0.04 \pm 0.02$ & $346.52 \pm 2.59$ & Yes \\
    HD 135344B & Outer & $14.80 \pm 0.90$ & $25.09 \pm 1.12$ & $86.21 \pm 0.51$ & $0.17 \pm 0.01$ & $26.49 \pm 2.48$ & No \\
    HD     139614 & Outer & $50.31 \pm 8.47$ & $25.79 \pm 2.24$ & $173.72 \pm 2.14$ & $0.29 \pm 0.05$ & $137.09 \pm 7.68$ & Yes \\
    HD     142527 & Outer & $47.28 \pm 1.75$ & $26.85 \pm 0.63$ & $174.19 \pm 0.53$ & $0.27 \pm 0.01$ & $10.35 \pm 2.44$ & Yes \\
    HD     163296 & Outer & $13.50 \pm 0.46$ & $45.96 \pm 0.58$ & $82.59 \pm 0.25$ & $0.16 \pm 0.01$ & $129.86 \pm 0.49$ & Yes \\
    HD     169142 & Inner & $1.28 \pm 0.07$ & $18.52 \pm 0.48$ & $35.27 \pm 0.07$ & $0.04 \pm 0.00$ & $242.80 \pm 2.32$ & Yes \\
     & Outer & $12.51 \pm 2.76$ & $19.61 \pm 1.15$ & $68.50 \pm 0.30$ & $0.18 \pm 0.04$ & $249.01 \pm 4.09$ & Yes \\
    HD     179218 & Outer & $5.65 \pm 3.87$ & $38.24 \pm 1.29$ & $123.13 \pm 1.22$ & $0.05 \pm 0.03$ & $201.09 \pm 2.36$ & No \\
    HD     290380 & Outer & $7.84 \pm 1.69$ & $25.25 \pm 4.62$ & $58.06 \pm 1.66$ & $0.14 \pm 0.03$ & $6.89 \pm 13.16$ & No \\
    HD     294260 & Outer & $34.12 \pm 2.97$ & $16.52 \pm 1.10$ & $133.74 \pm 0.56$ & $0.26 \pm 0.02$ & $8.33 \pm 84.85$ & Yes \\
    HD     31648 & Inner & $4.37 \pm 0.55$ & $34.46 \pm 1.14$ & $53.05 \pm 0.37$ & $0.08 \pm 0.01$ & $320.20 \pm 2.32$ & No \\
    HD     34282 & Inner & $20.75 \pm 1.42$ & $65.86 \pm 2.37$ & $175.03 \pm 1.08$ & $0.12 \pm 0.01$ & $293.52 \pm 0.23$ & Yes \\
     & Middle & $58.24 \pm 2.37$ & $60.97 \pm 1.93$ & $231.50 \pm 1.54$ & $0.25 \pm 0.01$ & $294.77 \pm 0.45$ & Yes \\
     & Outer & $112.68 \pm 9.97$ & $59.13 \pm 4.21$ & $327.37 \pm 4.99$ & $0.34 \pm 0.03$ & $297.03 \pm 1.32$ & Yes \\
    HD          34700 & Outer & $20.65 \pm 0.59$ & $40.89 \pm 0.32$ & $169.86 \pm 0.44$ & $0.12 \pm 0.00$ & $78.53 \pm 0.66$ & Yes \\
    HD          36112 & Outer & $10.73 \pm 0.32$ & $28.85 \pm 0.62$ & $79.23 \pm 0.19$ & $0.14 \pm 0.00$ & $178.50 \pm 2.13$ & No \\
     & Inner & $8.94 \pm 0.23$ & $23.79 \pm 0.50$ & $48.86 \pm 0.18$ & $0.18 \pm 0.01$ & $93.44 \pm 0.88$ & No \\
    HD          97048 & Inner & $8.89 \pm 0.82$ & $41.31 \pm 0.52$ & $116.47 \pm 0.44$ & $0.08 \pm 0.01$ & $11.94 \pm 1.13$ & Yes \\
     & Inner (2) & $31.51 \pm 1.53$ & $40.91 \pm 0.46$ & $196.59 \pm 0.70$ & $0.16 \pm 0.01$ & $6.99 \pm 0.74$ & Yes \\
     & Middle & $25.53 \pm 10.88$ & $44.17 \pm 2.11$ & $285.97 \pm 2.03$ & $0.09 \pm 0.04$ & $0.11 \pm 4.09$ & Yes \\
     & Outer & $56.36 \pm 8.39$ & $39.78 \pm 1.31$ & $372.27 \pm 2.68$ & $0.15 \pm 0.02$ & $357.08 \pm 2.83$ & Yes \\
    Hen     3-545 & Outer & $6.75 \pm 1.27$ & $33.82 \pm 1.76$ & $83.36 \pm 1.13$ & $0.08 \pm 0.01$ & $36.00 \pm 3.82$ & No \\
    HP     Cha & Inner & $8.48 \pm 2.31$ & $34.53 \pm 0.86$ & $68.77 \pm 0.46$ & $0.12 \pm 0.03$ & $131.26 \pm 2.24$ & Yes \\
    IM     Lup & Inner & $18.09 \pm 0.57$ & $53.87 \pm 1.00$ & $106.21 \pm 0.42$ & $0.17 \pm 0.01$ & $325.15 \pm 0.39$ & Yes \\
     & Inner (2) & $34.14 \pm 0.55$ & $49.28 \pm 0.40$ & $171.63 \pm 0.32$ & $0.20 \pm 0.00$ & $315.43 \pm 0.26$ & Yes \\
     & Middle & $56.50 \pm 4.75$ & $49.10 \pm 1.00$ & $243.73 \pm 2.25$ & $0.23 \pm 0.02$ & $327.66 \pm 1.64$ & Yes \\
     & Outer & $91.18 \pm 4.75$ & $50.58 \pm 1.94$ & $362.85 \pm 2.72$ & $0.25 \pm 0.01$ & $321.39 \pm 1.53$ & Yes \\
    IP     Tau & Outer & $9.03 \pm 1.48$ & $24.57 \pm 2.81$ & $39.87 \pm 0.53$ & $0.23 \pm 0.04$ & $166.32 \pm 7.24$ & Yes \\
    KK     Oph & Outer & $10.40 \pm 1.20$ & $45.06 \pm 2.30$ & $75.91 \pm 1.24$ & $0.14 \pm 0.02$ & $60.76 \pm 2.31$ & No \\
    PDS     66 & Middle & $12.12 \pm 1.55$ & $25.05 \pm 1.12$ & $75.92 \pm 0.29$ & $0.16 \pm 0.02$ & $193.85 \pm 3.95$ & Yes \\
    PDS     70 & Outer & $19.77 \pm 1.70$ & $54.62 \pm 2.06$ & $110.95 \pm 0.94$ & $0.18 \pm 0.01$ & $339.88 \pm 1.04$ & Yes \\
    RU     Lup & Inner & $6.18 \pm 1.32$ & $23.30 \pm 2.49$ & $61.51 \pm 0.65$ & $0.10 \pm 0.02$ & $142.46 \pm 9.29$ & Yes \\
     & Outer & $17.47 \pm 7.17$ & $20.37 \pm 2.85$ & $120.00 \pm 1.29$ & $0.15 \pm 0.06$ & $136.01 \pm 10.80$ & Yes \\
    RXJ1615-3255 & Inner & $4.30 \pm 0.72$ & $48.29 \pm 0.35$ & $69.82 \pm 0.21$ & $0.06 \pm 0.01$ & $144.99 \pm 0.36$ & No \\
     & Inner (2) & $27.30 \pm 3.38$ & $44.10 \pm 1.27$ & $145.76 \pm 1.18$ & $0.19 \pm 0.02$ & $146.72 \pm 1.32$ & No \\
     & Middle & $31.93 \pm 2.07$ & $46.45 \pm 0.92$ & $203.79 \pm 0.54$ & $0.16 \pm 0.01$ & $146.24 \pm 0.48$ & No \\
     & Outer & $53.29 \pm 2.35$ & $47.25 \pm 0.67$ & $277.49 \pm 1.62$ & $0.19 \pm 0.01$ & $148.26 \pm 3.01$ & No \\
    RY     Lupi & Outer & $16.52 \pm 1.07$ & $64.84 \pm 2.81$ & $102.79 \pm 0.78$ & $0.16 \pm 0.01$ & $107.25 \pm 0.43$ & Yes \\
    RY     Tau & Outer & $12.23 \pm 0.91$ & $66.72 \pm 3.46$ & $64.36 \pm 0.45$ & $0.19 \pm 0.01$ & $202.54 \pm 0.54$ & Yes \\
    SZ     111 & Outer & $25.72 \pm 2.38$ & $47.68 \pm 2.19$ & $104.99 \pm 1.21$ & $0.24 \pm 0.02$ & $225.90 \pm 1.82$ & Yes \\
    Sz     68 & Outer & $7.28 \pm 1.64$ & $37.46 \pm 2.16$ & $68.13 \pm 1.65$ & $0.11 \pm 0.02$ & $171.07 \pm 2.83$ (*) & Yes \\
    SZ     Cha & Middle & $12.72 \pm 0.35$ & $45.33 \pm 0.46$ & $77.95 \pm 0.18$ & $0.16 \pm 0.01$ & $340.30 \pm 0.38$ & Yes \\
     & Outer & $24.82 \pm 0.97$ & $45.44 \pm 0.94$ & $139.96 \pm 0.56$ & $0.18 \pm 0.01$ & $340.88 \pm 0.97$ & Yes \\
    TW     Hya & Inner & $0.84 \pm 0.12$ & $15.04 \pm 1.92$ & $21.92 \pm 0.17$ & $0.04 \pm 0.01$ & $58.17 \pm 7.91$ & No \\
     & Inner (2) & $1.73 \pm 0.36$ & $19.80 \pm 2.35$ & $36.23 \pm 0.29$ & $0.05 \pm 0.01$ & $42.05 \pm 8.80$ & No \\
     & Inner (3) & $3.68 \pm 0.58$ & $10.45 \pm 1.30$ & $48.24 \pm 0.17$ & $0.08 \pm 0.01$ & $63.34 \pm 12.53$ & No \\
     & Middle & $4.26 \pm 1.02$ & $10.51 \pm 1.97$ & $69.22 \pm 0.27$ & $0.06 \pm 0.01$ & $52.57 \pm 41.73$ & No \\
     & Outer & $15.03 \pm 7.01$ & $8.86 \pm 3.81$ & $110.33 \pm 0.58$ & $0.14 \pm 0.06$ & $339.07 \pm 31.38$ & No \\
     & Outer (2) & $21.94 \pm 7.04$ & $15.19 \pm 3.58$ & $158.48 \pm 1.43$ & $0.14 \pm 0.04$ & $22.19 \pm 18.56$ & No \\
    TYC     6213-1122-1 & Outer & $5.05 \pm 0.29$ & $62.22 \pm 0.31$ & $89.88 \pm 0.74$ & $0.06 \pm 0.00$ & $35.13 \pm 0.31$ & No \\
    T     Cha & Outer & $9.45 \pm 1.66$ & $60.11 \pm 2.17$ & $56.33 \pm 2.29$ & $0.17 \pm 0.03$ & $117.54 \pm 1.99$ & No \\
    UX     Tau     A & Outer & $8.07 \pm 1.62$ & $40.60 \pm 1.57$ & $89.85 \pm 0.62$ & $0.09 \pm 0.02$ & $166.27 \pm 1.92$ & Yes \\
    UZ     Tau & Inner & $3.92 \pm 0.56$ & $66.24 \pm 5.06$ & $54.34 \pm 0.70$ & $0.07 \pm 0.01$ & $257.22 \pm 44.01$ & Yes \\
    V1094     Sco & Outer & $6.54 \pm 1.72$ & $59.42 \pm 3.30$ & $64.79 \pm 0.70$ & $0.10 \pm 0.03$ & $108.04 \pm 0.76$ & Yes \\
    V1098     Sco & Outer & $11.13 \pm 1.61$ & $32.77 \pm 0.94$ & $102.60 \pm 0.82$ & $0.11 \pm 0.02$ & $14.28 \pm 2.39$ & No \\
    V1247     Ori & Outer & $34.45 \pm 1.37$ & $41.87 \pm 1.20$ & $152.40 \pm 1.08$ & $0.23 \pm 0.01$ & $111.06 \pm 1.32$ & Yes \\
    V351     Ori & Inner & $57.75 \pm 1.59$ & $66.57 \pm 0.28$ & $260.39 \pm 2.61$ & $0.22 \pm 0.01$ & $138.16 \pm 0.29$ & No \\
     & Outer & $95.72 \pm 3.97$ & $64.81 \pm 0.72$ & $369.83 \pm 5.49$ & $0.26 \pm 0.01$ & $140.91 \pm 1.05$ & No \\
    V4046     Sgr & Inner & $0.59 \pm 0.09$ & $31.64 \pm 1.12$ & $17.61 \pm 0.06$ & $0.03 \pm 0.01$ & $76.68 \pm 1.05$ & Yes \\
     & Middle & $0.49 \pm 0.06$ & $32.17 \pm 0.98$ & $31.90 \pm 0.07$ & $0.01 \pm 0.00$ & $75.23 \pm 0.95$ & Yes \\
     & Outer & $5.00 \pm 1.22$ & $35.89 \pm 1.69$ & $53.25 \pm 0.53$ & $0.09 \pm 0.02$ & $77.99 \pm 2.69$ & Yes \\
    V409     Tau & Outer & $7.65 \pm 0.85$ & $68.04 \pm 5.42$ & $61.18 \pm 0.71$ & $0.12 \pm 0.01$ & $44.71 \pm 0.72$ & Yes \\
    V599     Ori & Outer & $45.58 \pm 1.28$ & $59.45 \pm 0.49$ & $226.67 \pm 1.36$ & $0.20 \pm 0.01$ & $316.80 \pm 0.63$ & No \\
    V710     Tau & Outer & $10.65 \pm 0.70$ & $49.07 \pm 1.83$ & $51.74 \pm 0.43$ & $0.21 \pm 0.01$ & $264.03 \pm 1.04$ & Yes \\
    V721     CrA & Outer & $33.82 \pm 1.98$ & $61.85 \pm 0.81$ & $124.30 \pm 1.81$ & $0.27 \pm 0.02$ & $286.05 \pm 0.99$ & No \\
    V836     Tau & Outer & $4.72 \pm 0.86$ & $42.77 \pm 3.23$ & $44.25 \pm 0.80$ & $0.11 \pm 0.02$ & $115.11 \pm 3.47$ & Yes \\
    V935     Sco & Outer & $8.72 \pm 0.57$ & $48.25 \pm 2.44$ & $54.19 \pm 0.72$ & $0.16 \pm 0.01$ & $83.92 \pm 14.66$ & Yes \\ 
    WISPIT 2 (H-Band) & Inner & $1.70 \pm 0.15$ & $44.95 \pm 0.39$ & $38.44 \pm 0.08$ & $0.04 \pm 0.00$ & $1.54 \pm 0.28$ & No \\
     & Inner (2) & $10.44 \pm 0.62$ & $41.81 \pm 0.56$ & $96.73 \pm 0.63$ & $0.11 \pm 0.01$ & $357.10 \pm 1.55$ & No \\
     & Middle & $24.05 \pm 1.15$ & $45.40 \pm 1.07$ & $163.54 \pm 2.93$ & $0.15 \pm 0.01$ & $356.51 \pm 1.31$ & No \\
     & Outer & $76.08 \pm 6.32$ & $43.62 \pm 1.54$ & $316.36 \pm 4.50$ & $0.24 \pm 0.02$ & $357.81 \pm 1.56$ & No \\
    WISPIT 2 (K-Band) & Inner & $1.09 \pm 0.48$ & $43.97 \pm 0.88$ & $37.22 \pm 0.12$ & $0.03 \pm 0.01$ & $0.243 \pm 0.48$ & No \\
     & Inner (2) & $6.62 \pm 1.58$ & $42.64 \pm 1.27$ & $102.75 \pm 1.36$ & $0.06 \pm 0.01$ & $359.34 \pm 4.46$ & No \\
     & Middle & $17.93 \pm 2.73$ & $47.52 \pm 1.27$ & $156.30 \pm 1.76$ & $0.12 \pm 0.02$ & $356.82 \pm 1.50$ & No \\
    WRAY     15-1384 & Inner & $0.53 \pm 0.21$ & $58.27 \pm 0.96$ & $70.63 \pm 1.28$ & $0.01 \pm 0.00$ & $53.72 \pm 1.21$ & No \\
    WRAY     15-1400 & Inner & $6.34 \pm 0.94$ & $41.32 \pm 2.08$ & $66.31 \pm 1.27$ & $0.10 \pm 0.01$ & $330.44 \pm 2.86$ & No \\
    WRAY     15-1880 & Inner & $6.74 \pm 0.73$ & $41.94 \pm 1.38$ & $56.46 \pm 0.88$ & $0.12 \pm 0.01$ & $221.00 \pm 2.88$ & No \\
    WRAY     15-788 & Inner (2) & $6.30 \pm 1.78$ & $26.25 \pm 5.05$ & $56.59 \pm 0.55$ & $0.11 \pm 0.03$ & $57.56 \pm 58.14$ & No \\
     & Middle & $11.92 \pm 1.83$ & $27.00 \pm 1.13$ & $98.82 \pm 0.67$ & $0.12 \pm 0.02$ & $40.50 \pm 3.18$ & No \\
     & Outer & $13.79 \pm 16.82$ & $31.97 \pm 9.13$ & $166.00 \pm 5.88$ & $0.08 \pm 0.10$ & $61.21 \pm 10.07$ & No \\
    WW     Cha & Inner & $7.98 \pm 3.32$ & $45.36 \pm 3.56$ & $63.29 \pm 0.53$ & $0.13 \pm 0.05$ & $31.25 \pm 4.97$ & Yes \\
     & Outer & $18.84 \pm 0.67$ & $43.58 \pm 0.77$ & $118.23 \pm 0.43$ & $0.16 \pm 0.01$ & $34.35 \pm 1.50$ & Yes \\
    \end{longtable}

    \section{Figure of Ellipse Fits for all Disks}
    \label{AllDiskFigure}

    Shown below is the figure containing all the ellipse fits derived through this work, overlaid on their respective image. The ellipse fit shown is the constrained fit if available and the free fit otherwise. The disks are sorted into their respective disk morphologies defined in Sect. \ref{Global Results} for comparison.

    \begin{figure*}
        \centering
        \includegraphics[width=0.99\textwidth]{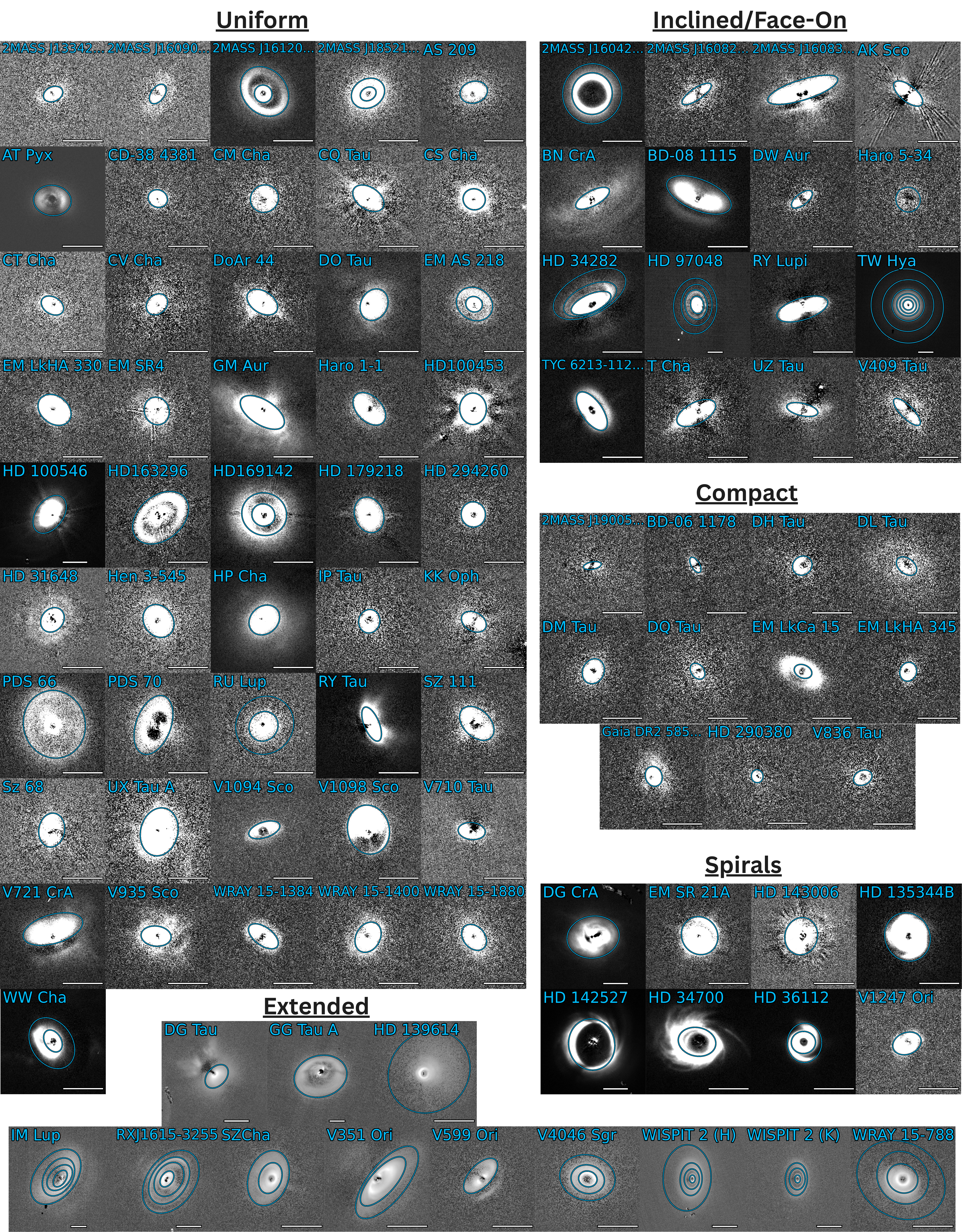}
        \caption{Ellipse fits described in Sect.~\ref{Ellipse Fitting} overlaid on all 92 unique disks, categorised by the disk type defined in Sect. \ref{Global Results}. Scattered-light images are displayed using either a logarithmic stretch or a linear scale with fixed $v_{\min}$ and $v_{\max}$ to optimise visibility of disk structure. The scale bar in each panel denotes 1 arcsecond.}
        \label{fig:2ndstepoutlierremoval}
    \end{figure*}

    \clearpage
    \newpage

    \section{Outlier Removal}
    \label{outlierremoval}

    Due to the faint and noisy nature of the images, the detected points are often irregular and inconsistent, with spurious detections occurring across closely spaced radial passes. Several stages of outlier removal are implemented within the algorithm to combat clear outliers, hereafter also referred to as false detections.

    First, extreme outliers are removed by computing the radial distance of each detected point from the image centre (i.e. the coronagraph position) and discarding those exceeding the mean radial distance of all points by a user-defined threshold (typically $\ge$50 pixels), which changes depending on whether one is extracting information from a feature close (smaller threshold) or distant from the centre (larger threshold).
    
\begin{figure}[h!]
    \centering
    \includegraphics[width=0.49\textwidth]{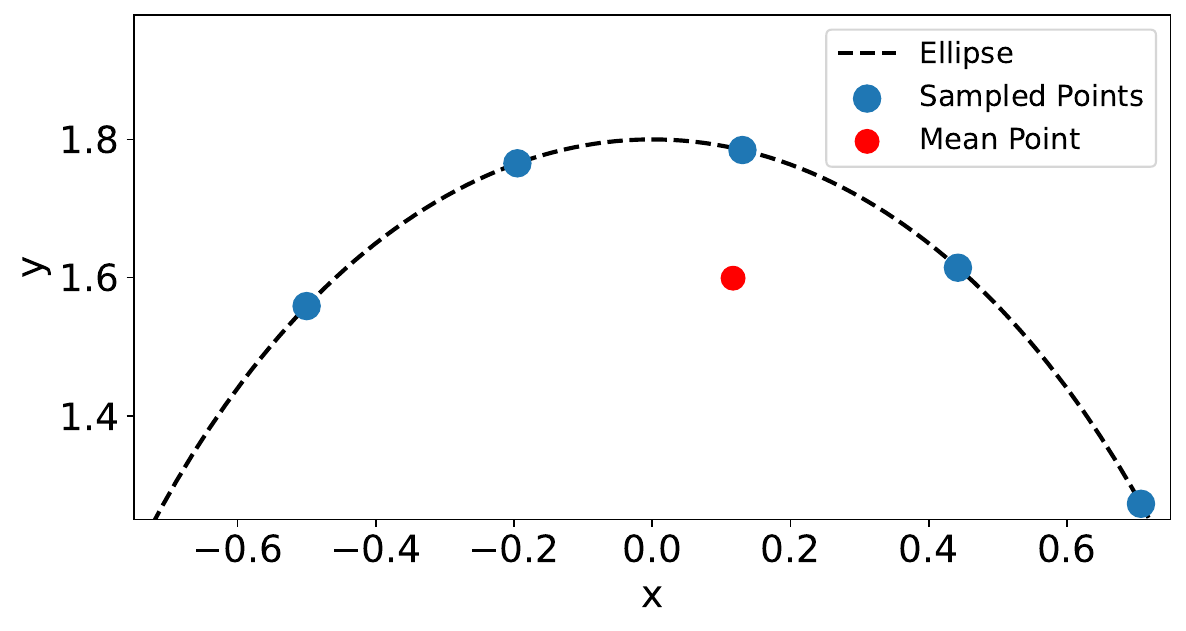}
    \caption{Illustration of point binning with excessive angular bin size. When the bin angle is too large, true edge points (blue) are truncated, and the resulting averaged point (red) no longer accurately represents the underlying distribution.}
    \label{fig:BinnedIssue}
    \end{figure}

    Points are then spatially binned based on azimuthal angle. High S/N disks with well-defined edges require little to no binning, while poorly defined or noisy edges benefit from binning to suppress spurious structure. Binning reduces the total number of data points and smooths out local irregularities, however may interfere with other outlier removal processes and their user-defined parameters must be adjusted to account for the decrease in points and increase in distance between them. Therefore, the bin angle selected varies between disks, with little/no binning for extended structures and greater degrees of binning for compact, rough-edged disks. It is noted that one must consider that using a large binning angle ($\ge5^\circ$) causes the averaged coordinates to systematically truncate inward, since the binned points approximate a curved structure, as illustrated in Fig.~\ref{fig:BinnedIssue}.

    An outlier removal step is now performed within each bin by discarding points exceeding the bin's mean radial distance by a smaller threshold (typically 10 pixels), aiming to remove extreme outliers from their local bin. After averaging each bin, a final pass removes bins whose mean radius deviates from the total mean of all bins by more than 20-30 pixels. This helps eliminate bins comprised of entirely false detections but can be relaxed for highly eccentric structures, where the disk edge greatly varies in distance from the star azimuthally.

    With global and binned outliers removed, to identify and remove local outliers that deviate from the structure's natural curve along the extracted azimuthal profile, a two-stage filtering process was implemented, visualised entirely in Fig.~\ref{fig:2ndstepoutlierremoval}. 
    
\begin{figure}[h]
        \centering
        \includegraphics[width=1\linewidth]{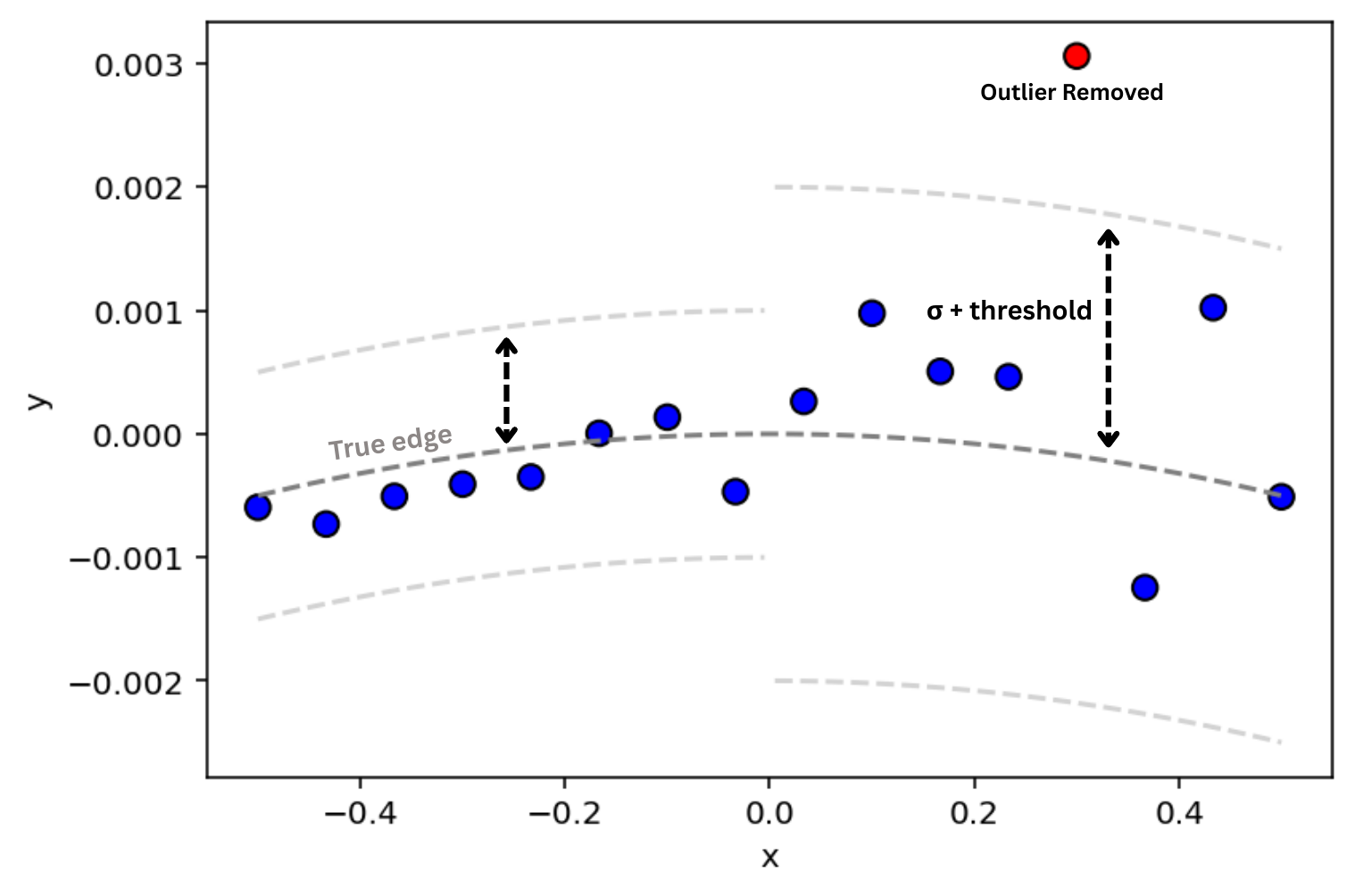}
        \caption{Diagram illustrating the first step of local outlier removal, where points are identified as outliers if their radial distance exceeds the standard deviation of the previous ten points plus a defined threshold.}
        \label{fig:outlieriterative}
    \end{figure}

    In the first stage (shown in Fig.~\ref{fig:outlieriterative} and together with stage 2 in Fig.~\ref{fig:2ndstepoutlierremoval}), data points were scanned sequentially using a sliding-window technique. Each point was compared to the local mean radial distance, obtained from the trimmed mean of the previous ten points (central eight by value) to reduce outlier influence. Naturally, the first ten points were excluded from this removal step (see Fig.~\ref{fig:2ndstepoutlierremoval}); therefore the image must be rotated to an angle where the edge is smooth and well-defined (e.g, the forward-scattering side). Points deviating beyond the trimmed mean plus a fixed threshold (10 pixels was used in this work) were identified as false detections. To handle structural discontinuities (e.g., gaps or disk breaks due to shadowing or faint signal on the back-scattering side), a new section was initiated once the number of consecutive outliers exceeded a predefined ‘bridge’ threshold (typically 20 rejections), preventing over-rejection near features like the noisy back-scattering side.
    
    In the second stage of the local outlier removal (shown alongside stage 1 in Fig.~\ref{fig:2ndstepoutlierremoval}), each section was independently processed using 1.5$\sigma$ clipping around its mean radial distance to remove residual outliers. This targets clusters of false detections that passed initial filtering, especially in low S/N regions where noise near the disk edge can bias binned averages outward. These biased bins inflate local standard deviations, reducing the effectiveness of earlier filtering. To prevent over-rejection, the first (typically largest) section on the high S/N front-scattering side is exempt, as strict clipping would wrongly exclude points near the curve ends due to natural curvature.

    Outside of the previously mentioned, predominately automatic outlier removal steps, the algorithm allows for manual exclusion of bins within specific angular ranges. These user-defined removals can be applied either before or after the outlier rejection process. When applied beforehand, they prevent problematic regions, such as regions of consistently poor signal (e.g. the back-scattering side), from influencing the initial threshold-based or rolling average calculations. When applied after outlier rejection, they allow for targeted clean-up of residual artefacts, particularly in edge cases where structured noise or imaging artefacts remain.

    \begin{figure*}[t!]
    \centering
    \sidecaption
        \includegraphics[width=0.7\textwidth]{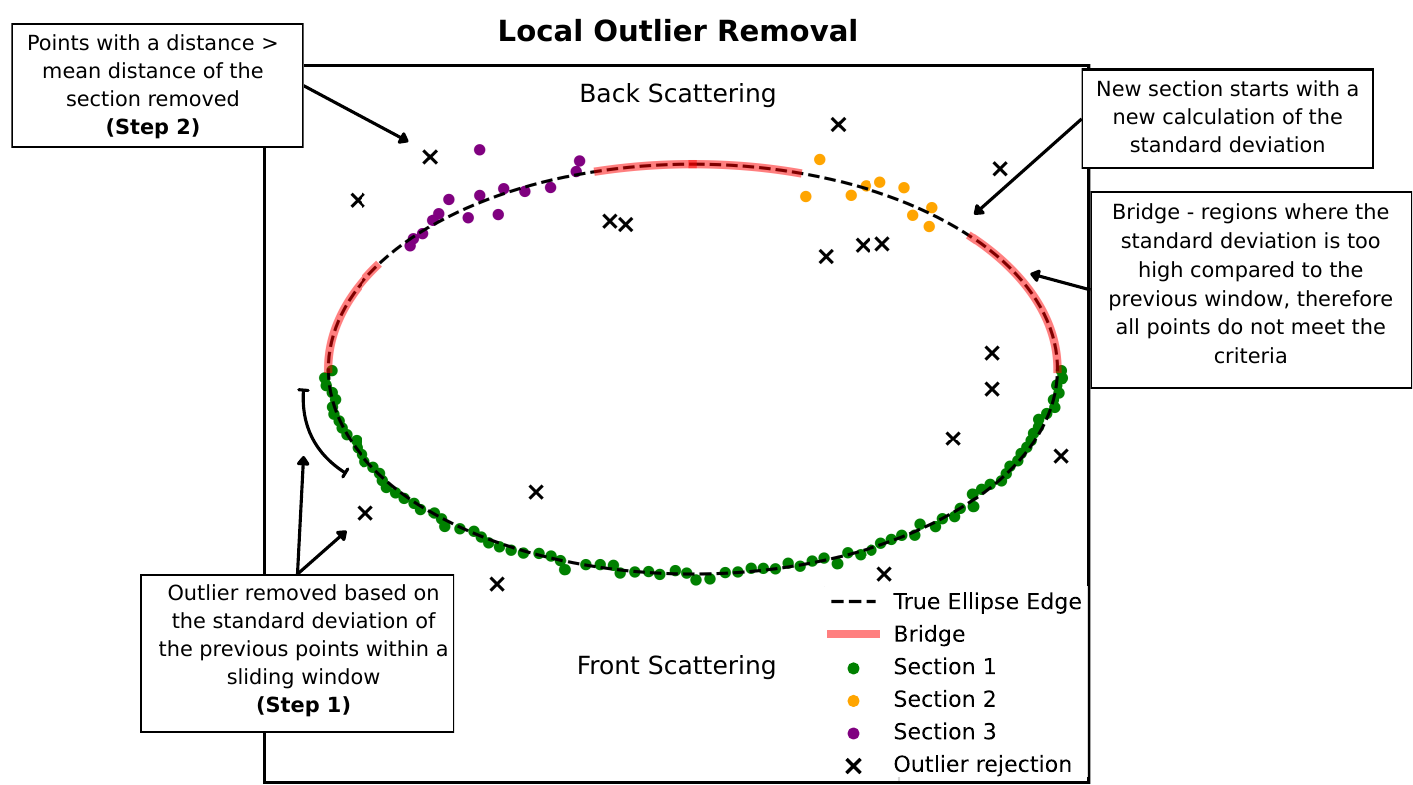}
        \caption{Diagram illustrating the two-step local outlier removal process. The elliptical disk is divided into angular sections based on series of outlier rejections from step 1. Section 1 (green) spans the bright, front-scattering side and undergoes step 1 filtering. Sections 2 (yellow) and 3 (purple) lie on the fainter, back-scattering side and are targeted by the second-stage rejection (black crosses). Red arcs (a 'bridge') mark angular gaps where the local standard deviation is inflated due to lower S/N, preventing the sliding window from identifying outliers. Defining new sections in these regions enables localised standard deviation recalculation, allowing for successful outlier detection across the back-scattering side. }
        \label{fig:2ndstepoutlierremoval}
    \end{figure*}

    \section{Coverage and PA Convention}
    \label{Coverage and PA convention}

    \begin{figure}[b!]
        \centering
        \includegraphics[width=1\linewidth]{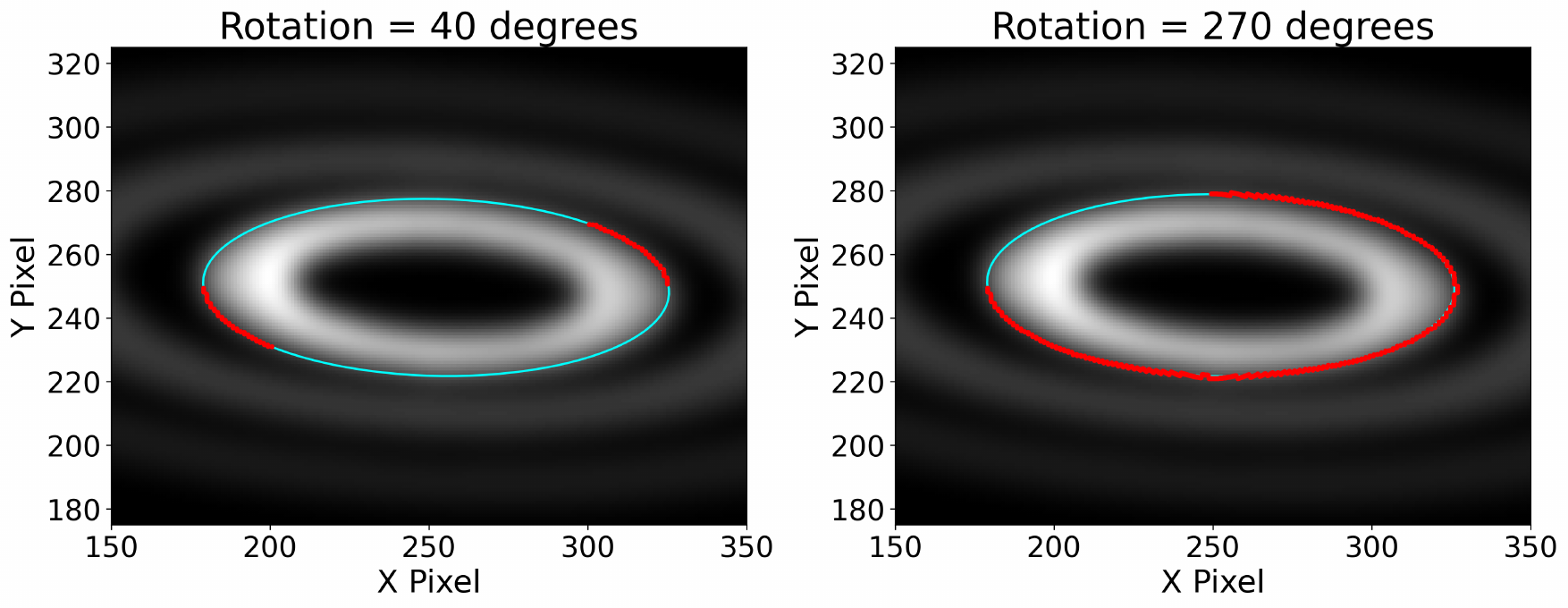}
        \caption{Example of two free ellipse fits (cyan) applied to the same model disk, where edge points (red) were extracted at increasing azimuthal angles in one direction. \textbf{Left:} the detected edge points span 270$^\circ$ of a single side of the ring. \textbf{Right:} points span only 40$^\circ$ but are located on opposing sides of the ring. Both ellipses accurately map the underlying structure, showing the importance of extracting information from various azimuthal regions of the disk.
    }
        \label{fig:rottest}
    \end{figure}

    To evaluate how azimuthal coverage of the extracted points from the structure extraction step of the algorithm impacts the accuracy of free ellipse fit, two configurations were tested on the same model disk, shown in Fig.~\ref{fig:rottest}. In each case, edge points were incrementally added until the ellipse solution converged. For the first configuration, convergence occurred after accumulating 270$^\circ$ of continuous coverage on one side of the disk. In the second configuration, convergence was achieved with only 40$^\circ$ of total azimuthal coverage, where the points were distributed equally on opposing sides of the ring. This result demonstrates that accurate ellipse fitting depends more critically on azimuthal distribution than on total angular span. Even limited angular coverage can constrain the fit effectively, providing the points sample diverse regions of the disk geometry.

    \begin{figure}[h]
        \centering
        \includegraphics[width=1\linewidth]{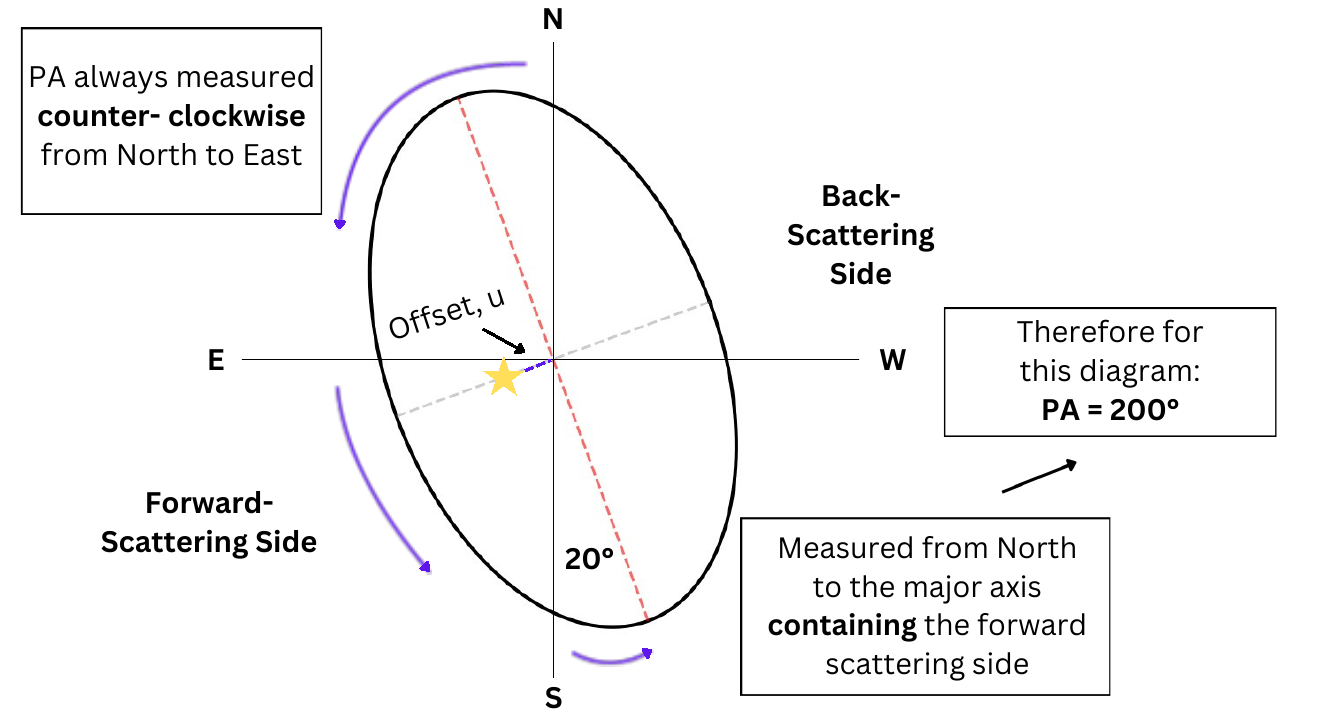}
        \caption{Diagram showing the PA convention used in this work. The PA is measured North to East, counter-clockwise and to the major-axis containing the forward-scattering region of the disk.
    }
        \label{fig:PAeg}
    \end{figure}

    The PA obtained directly from the ellipse fit is measured counter-clockwise from the positive x-axis to the first instance of the major axis. To make this consistent with the typical convention, which is illustrated in Fig. \ref{fig:PAeg} and used throughout this work. The typical convention for PA is measured from North through East to the first instance of the major axis such that the forward-scattering side lies between this axis and its opposite counterpart. To correct the PA measured from the ellipse fit, an offset of 90$^\circ$ was added to convert the measurement to start from the +x axis to +y axis. If this angle did not contain the forward scattering side, then an additional offset of 180$^\circ$ was added, giving offset options of 90$^\circ$ and 270$^\circ$. The forward-scattering side was identified by comparing phase functions derived for each offset, inspecting the corresponding disk-region overlay fits, and assessing the global disk geometry and brightness distribution for each side of the disk to determine which side is the forward and back scattering region (further discussed in Coyne et al., in prep.).

    When the forward-scattering side was ambiguous, a vector was drawn from the stellar position to the centre of the fitted ellipse; for an inclined disk this vector should point toward the back-scattering side, since the star appears offset toward the forward-scattering side in the image plane. In some cases, the vector lay approximately along the major axis and was therefore unreliable. For these systems, the reported PA is defined simply as the angle from North to the first occurrence of the major axis, independent of any forward-scattering determination.

    \section{Wavelength Dependence Analysis}
    \label{Wavelength Dependence Analysis}

\begin{figure}[h!]
        \centering
        \includegraphics[width=1\linewidth]{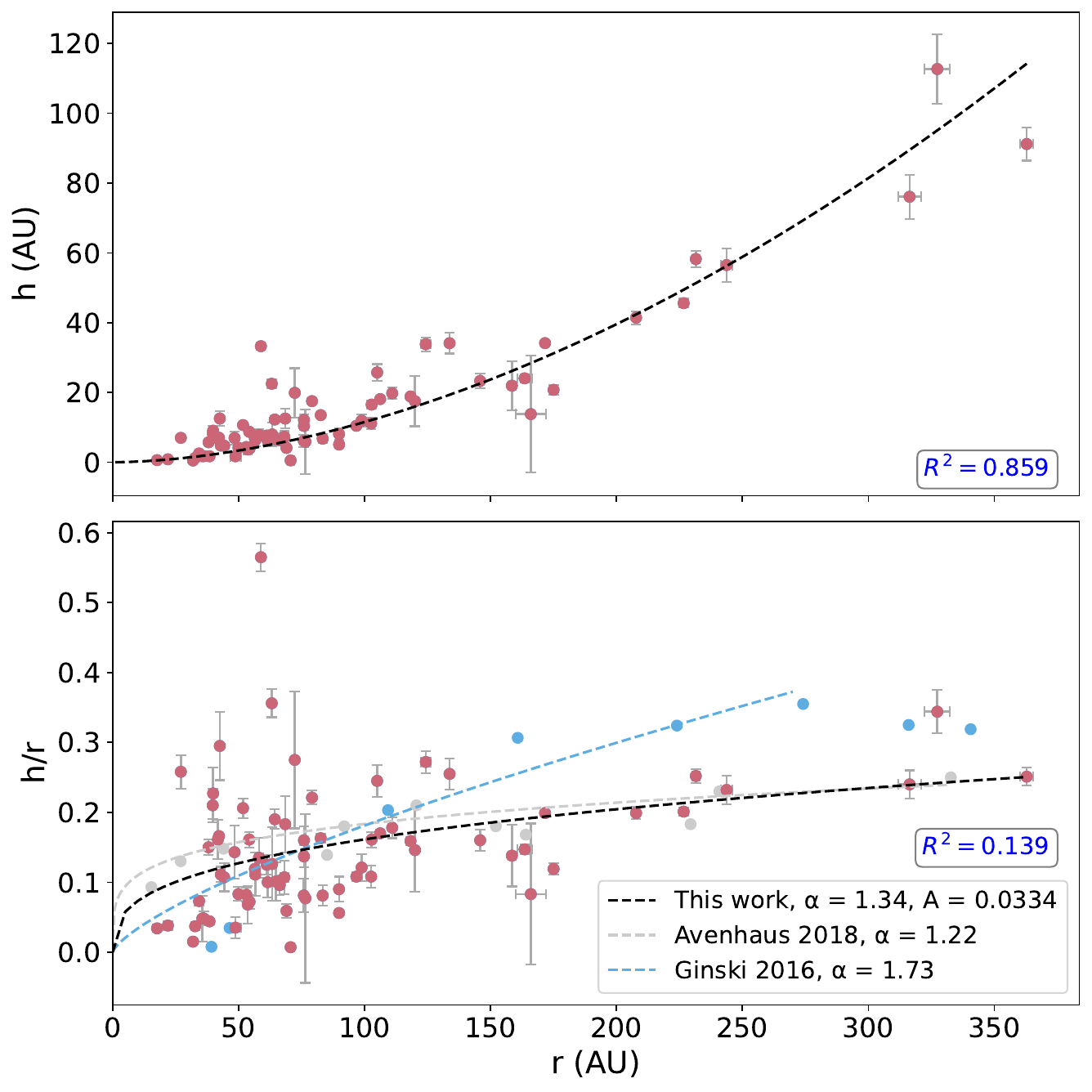}
        \caption{$h$ (\textbf{Top}) and $h/r$ (\textbf{Bottom}) versus separation ($r$) for the restricted dataset, excluding spiral and lowly or highly inclined disks and now including only H-band observations. As the majority of the dataset consists of H-band images, the resulting distributions and fitted trends show minimal deviation from the restricted dataset containing all filters.
        }
        \label{fig:onlyH}
    \end{figure}

    The dataset includes observations in the J (1.1 - 1.4 $\mu$m), H (1.5 - 1.8 $\mu$m) and K (2.0 - 2.4 $\mu$m) bands, each sampling slightly different heights in the disk surface. To assess the impact of this wavelength dependence on the derived aspect ratios, the analysis was repeated using only the H-band subset, with the previously identified difficult-to-fit groups removed (see Sect. \ref{Global Results}), as shown in Fig.~\ref{fig:onlyH} (totalling 40 unique disks and 58 measurements). Since H-band observations constitute the majority of the dataset, the resulting correlations remained mostly consistent, suggesting that band-dependent effects do not significantly influence the overall trend. The coefficient of determination \( R^2 \) decreased from 0.166 to 0.139, indicating a slight, but not significant, decrease in correlation when limiting the sample. This decrease contrasts with the expected improvement following dataset standardisation, although the decrease is minor.
    
    As seen from the analysis on WISPIT~2 (Sect. \ref{Multi-Ringed Systems & WISPIT 2}), wavelength dependency does indeed play a significant role for individual systems imaged in different filters and would also influence datasets with a more evenly spread filter distribution. However, since the dataset contains predominately H-band images, the effect does not substantially contribute to the scatter observed in the $h/r$ versus $r$ relations presented here.

    \section{Signal Excesses Found in Multi-ringed Systems}
    \label{Signal Excesses}

    \begin{figure*}[t]
        \centering
        \includegraphics[width=0.98\linewidth]{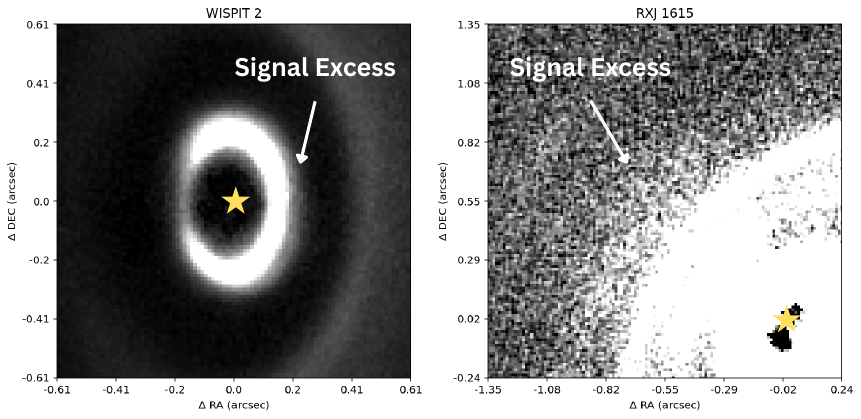}
        \caption{WISPIT~2 (\textbf{left}) and RXJ~1615 (\textbf{right}) with the excess signal annotated for each image. Both signal excesses appear along the minor axis and on the forward scattering side of the disk.}
        \label{fig:WISPvsRXJexcess}
    \end{figure*}

    For three multi-ringed systems RXJ~1615, HD~97048 and WISPIT~2, an excess emission has been observed along the minor axis. It is found within the forward scattering region for RXJ~1615 and WISPIT~2 and the back scattering region for HD~97048. The excess signal for WISPIT~2 is discussed in Sect. \ref{Multi-Ringed Systems & WISPIT 2} and in detail by \citet{VanCapelleveen2025}. To the best of knowledge, there has been no previous discussion of the signal excess within RXJ~1615.

    For HD~97048, \citet{Ginski_2016} noted excess emission along the eastern (back-scattering) side of Ring~2, which they attributed to a partially illuminated inner wall of the structure. Their fitting technique, which targets the peak flux, is largely insensitive to this extended signal at the ring’s edge. In contrast, this work employs an edge-detection method that traces the outer boundary of the emission (Gaussian fitting was not used for Ring 2 due to a lack of retained points during extraction), thereby including the excess flux in the fit. If this extended emission indeed originates from a vertical wall rather than the true disk signal, the resulting ellipse fit, offset, and inferred vertical profile may not reflect the true geometry of the ring. \citet{Rich2021} investigated this scenario by generating scattered light radiative transfer models using TORUS \citep{harries2019torusradiationtransfercode} for both single-ring and two-ring disk geometries, with radial extents comparable to the inner disk and Ring~2 of HD~97048. In the single-ring case, the ring is fully illuminated, producing a bright wall-like feature. In contrast, the two-ring configuration (as observed in HD~97048) experiences shadowing from the inner disk, and no such wall is observed. Following this, \citet{Rich2021} fitted an ellipse to Ring 2 with $h = 32.3 \pm 1.5$ au at $r = 172 \pm 4$ au, which results in $h/r = 0.188$, a higher flaring than this work ($h/r = 0.160$) but significantly lower than the aspect ratio reported by \citet{Ginski_2016} ($h/r = 0.307$).

\section{Stellar Mass and Vertical Height Correlation Derivation}
\label{StellarMassCalc}

Further to Sect. \ref{System Properties affecting the Vertical Height Structure}, we explicitly derive the expected dependence of the disk vertical scale height on stellar mass to demonstrate that if luminosity is not proportional to $M^4$, that the cancellation between the sound speed and Keplerian frequency no longer holds. In this case, the disk scale height acquires an explicit dependence on stellar mass through the mass-luminosity relation, such that variations in stellar luminosity at fixed mass, as expected for pre–main-sequence stars, may introduce trends in the disk aspect ratio.

The vertical scale height of a disk in hydrostatic equilibrium is given by
\begin{equation}
h = \frac{c_s}{\Omega_K},
\end{equation}
where $c_s$ is the isothermal sound speed and $\Omega_K$ is the Keplerian angular frequency. The Keplerian frequency scales as
\begin{equation}
\Omega_K = \sqrt{\frac{G M_*}{r^3}} \propto M_*^{1/2},
\end{equation}
at fixed radius. The sound speed is given by
\begin{equation}
c_s = \sqrt{\frac{k_B T}{\mu m_H}} \propto T^{1/2}.
\end{equation}
Assuming the disk temperature is dominated by stellar irradiation,
\begin{equation}
T \propto L_*^{1/4}.
\end{equation}
For a general mass--luminosity relation of the form
\begin{equation}
L_* \propto M_*^{\beta},
\end{equation}
the sound speed scales as
\begin{equation}
c_s \propto M_*^{\beta/8}.
\end{equation}
Combining these relations yields
\begin{equation}
h \propto \frac{c_s}{\Omega_K} \propto M_*^{(\beta - 4)/8}.
\end{equation}
For the classical main-sequence case where $\beta = 4$, the stellar-mass dependence cancels exactly. For pre--main-sequence stars, where $\beta$ is not fixed, an explicit dependence of the disk scale height on stellar mass is expected.

\section{Planet Opening Gaps Table}
\label{Planet Opening Gaps Table}

    Table \ref{GapTable} summarises the inferred planet masses associated with observed gaps in multi-ringed disks within the sample, based on the gap-opening model of \citet{Kanagawa2016} and as discussed in Sect. \ref{Planets Opening Gaps}. The gap width is measured as the midpoint between the two adjacent rings/disk, where the point is measured at the peak of the Gaussian or at the edge of a disk, depending on whether Gaussian fitting or edge detection was used. While the two methods measure different points within their respective structure, the signal (and distance) added or lost from the difference in the location of the points is negligible compared to the total gap width. $R_p$ denotes the separation of the inferred planet (e.g., the midpoint of the gap), the aspect ratio was calculated using the scale parameter \(A\) and flaring index \(\alpha\) derived from the power-law fits for individual systems in Sect. \ref{Multi-Ringed Systems & WISPIT 2} and where measurements were not made, the extended disk trend's parameters (see Sect. \ref{Global Results}) were used instead. Finally, the planet masses are listed for the three considered viscosity parameters.

    \begin{table*}[h!]
    \centering
    \begin{tabular}{lcccccc}
    \hline
    \shortstack{\textbf{System} \\ \relax} & \shortstack{Gap Width \\ (au)} & \shortstack{$R_p$ \\ (au)} & \shortstack{$h_p / r_p$ \\ \relax} & \shortstack{$M_p$ ($M_J$)\\ $(\alpha_{\mathrm{visc}}=10^{-2})$} & \shortstack{$M_p$ ($M_J$)\\ $(\alpha_{\mathrm{visc}}=10^{-3})$} & \shortstack{$M_p$ ($M_J$)\\ $(\alpha_{\mathrm{visc}}=10^{-4})$} \\
    \hline
    \textbf{J16120668-3010270} & $61.4 \pm 0.6$ & $61.1 \pm 0.3$ & 0.019 & $1.11 \pm 1.13$ & $0.35 \pm 0.36$ & $0.11 \pm 0.11$ \\
    \hline
    \textbf{J18521730-3700119} & $34.3 \pm 0.5$ & $48.5 \pm 0.2$ & 0.016 & $0.61 \pm 0.73$ & $0.19 \pm 0.23$ & $0.06 \pm 0.07$ \\
    \hline
    \textbf{HD  97048} & $80.2 \pm 1.2$ & $156.2 \pm 0.6$ & 0.026 & $1.73 \pm 1.26$ & $0.55 \pm 0.40$ & $0.17 \pm 0.13$ \\
     & $75.0 \pm 6.7$ & $233.8 \pm 3.3$ & 0.030 & $0.83 \pm 0.55$ & $0.26 \pm 0.17$ & $0.08 \pm 0.05$ \\
     & $97.1 \pm 10.0$ & $319.9 \pm 5.0$ & 0.033 & $0.87 \pm 0.53$ & $0.28 \pm 0.17$ & $0.09 \pm 0.05$ \\
    \hline
    \textbf{HD 169142} & $33.1 \pm 0.4$ & $51.8 \pm 0.2$ & 0.016 & $0.89 \pm 1.02$ & $0.28 \pm 0.32$ & $0.09 \pm 0.10$ \\
    \hline
    \textbf{HD 34282} & $58.5 \pm 2.2$ & $201.4 \pm 1.1$ & 0.044 & $0.76 \pm 0.34$ & $0.24 \pm 0.11$ & $0.08 \pm 0.03$ \\
     & $86.3 \pm 5.2$ & $273.8 \pm 2.6$ & 0.054 & $1.25 \pm 0.47$ & $0.39 \pm 0.15$ & $0.12 \pm 0.05$ \\
    \hline
    \textbf{IM Lup} & $65.8 \pm 0.5$ & $138.6 \pm 0.2$ & 0.047 & $1.42 \pm 0.59$ & $0.45 \pm 0.19$ & $0.14 \pm 0.06$ \\
     & $61.8 \pm 2.7$ & $202. 4 \pm 1.4$ & 0.053 & $0.71 \pm 0.27$ & $0.22 \pm 0.08$ & $0.07 \pm 0.03$ \\
     & $121.7 \pm 4.8$ & $294.2 \pm 2.4$ & 0.060 & $1.55 \pm 0.53$ & $0.49 \pm 0.17$ & $0.15 \pm 0.05$ \\
    \hline
    \textbf{RXJ1615-3255} & $75.9 \pm 1.2$ & $107.8 \pm 0.6$ & 0.023 & $1.17 \pm 0.97$ & $0.37 \pm 0.31$ & $0.12 \pm 0.10$ \\
     & $58.0 \pm 1.3$ & $174.8 \pm 0.6$ & 0.034 & $0.47 \pm 0.26$ & $0.15 \pm 0.08$ & $0.05 \pm 0.03$ \\
     & $73.7 \pm 1.7$ & $240.6 \pm 0.9$ & 0.043 & $0.58 \pm 0.26$ & $0.18 \pm 0.08$ & $0.06 \pm 0.03$ \\
    \hline
    \textbf{TW Hya} & $14.3 \pm 0.3$ & $29.1 \pm 0.2$ & 0.011 & $0.14 \pm 0.24$ & $0.05 \pm 0.08$ & $0.01 \pm 0.02$ \\
     & $12.0 \pm 0.3$ & $42.2 \pm 0.2$ & 0.015 & $0.07 \pm 0.09$ & $0.02 \pm 0.03$ & $0.01 \pm 0.01$ \\
     & $21.0 \pm 0.3$ & $58.7 \pm 0.2$ & 0.018 & $0.16 \pm 0.16$ & $0.05 \pm 0.05$ & $0.02 \pm 0.02$ \\
     & $41.1 \pm 0.6$ & $89.8 \pm 0.3$ & 0.025 & $0.40 \pm 0.31$ & $0.13 \pm 0.10$ & $0.04 \pm 0.03$ \\
     & $48.2 \pm 1.5$ & $134.4 \pm 0.8$ & 0.033 & $0.38 \pm 0.22$ & $0.12 \pm 0.07$ & $0.04 \pm 0.02$ \\
    \hline
    \textbf{V4046 Sgr} & $14.3 \pm 0.1$ & $24.7 \pm 0.0$ & 0.004 & $0.11 \pm 0.46$ & $0.03 \pm 0.15$ & $0.01 \pm 0.05$ \\
     & $20.8 \pm 0.6$ & $42.3 \pm 0.3$ & 0.014 & $0.44 \pm 0.60$ & $0.14 \pm 0.19$ & $0.04 \pm 0.06$ \\
    \hline
    \textbf{WISPIT 2 (H-band)} & $58.3 \pm 0.6$ & $67.6 \pm 0.3$ & 0.018 & $1.20 \pm 1.26$ & $0.38 \pm 0.40$ & $0.12 \pm 0.13$ \\
     & $66.8 \pm 3.0$ & $130.1 \pm 1.5$ & 0.030 & $0.93 \pm 0.59$ & $0.29 \pm 0.19$ & $0.09 \pm 0.06$ \\
     & $152.8 \pm 5.4$ & $240.0 \pm 2.7$ & 0.049 & $2.94 \pm 1.19$ & $0.93 \pm 0.38$ & $0.29 \pm 0.12$ \\
    \hline
    \textbf{WISPIT 2 (K-band)} & $65.5 \pm 1.4$ & $70.0 \pm 0.7$ & 0.011 & $0.66 \pm 1.15$ & $0.21 \pm 0.36$ & $0.07 \pm 0.12$ \\
     & $53.6 \pm 2.2$ & $129.5 \pm 1.1$ & 0.023 & $0.39 \pm 0.33$ & $0.12 \pm 0.10$ & $0.04 \pm 0.03$ \\
    \hline
    \textbf{WRAY 15-788} & $42.2 \pm 0.9$ & $77.7 \pm 0.4$ & 0.022 & $0.76 \pm 0.65$ & $0.24 \pm 0.21$ & $0.08 \pm 0.07$ \\
     & $67.2 \pm 5.9$ & $132.4 \pm 3.0$ & 0.032 & $1.17 \pm 0.72$ & $0.37 \pm 0.23$ & $0.12 \pm 0.07$ \\
    \hline
    \end{tabular}
    \caption{Estimated planet masses for each observed gap using the \citet{Kanagawa2016} model. The gap locations, planet semi-major axes ($R_p$) and aspect ratios are calculated based on the fitting procedure in this work. Planet masses (in Jupiter masses) are computed assuming three different values of the disk viscosity parameter $\alpha_{\mathrm{visc}}$.}
    \label{GapTable}
    \end{table*}

\end{appendix}
    
\end{document}